\documentclass[lettersize,journal]{IEEEtran}
\usepackage{amsmath,amsfonts}
\usepackage{algorithmic}
\usepackage{algorithm}
\usepackage{array}
\usepackage[caption=false,font=normalsize,labelfont=sf,textfont=sf]{subfig}
\usepackage{textcomp}
\usepackage{stfloats}
\usepackage{url}
\usepackage{verbatim}
\usepackage{graphicx}
\usepackage{adjustbox}
\usepackage{cite}
\usepackage{bm, amssymb, fixmath,dsfont}
\let\oldnl\nl
\newcommand{\nonl}{\renewcommand{\nl}{\let\nl\oldnl}}
\usepackage{color}
\usepackage{soul}
\usepackage{multirow}
\usepackage{booktabs}
\usepackage{bbm}
\usepackage{tikz}
\usepackage{rotating}

\newtheorem{definition}{Definition}

\usepackage{hyperref}
\hypersetup{colorlinks,breaklinks,linkcolor=blue,urlcolor=blue,anchorcolor=blue,citecolor=blue}

\begin{document}
	
	\title{A Value Based Parallel Update MCTS Method for Multi-Agent Cooperative Decision Making of Connected and Automated Vehicles}
	
	\author{
		Ye Han,~\IEEEmembership{Student Member,~IEEE,} Lijun Zhang$^*$,~\IEEEmembership{Member,~IEEE,} Dejian Meng ,~\IEEEmembership{Member,~IEEE,} Zhuang~Zhang,~\IEEEmembership{Student Member,~IEEE,} Xingyu Hu, and Songyu Weng
		\thanks{This paper was produced by the IEEE Publication Technology Group. They are in Piscataway, NJ.}
		\thanks{Manuscript received Mon Day, 2024; revised Mon Day, 2024.}
		\thanks{Corresponding author: Lijun Zhang (tjedu\_zhanglijun@tongji.edu.cn).}
		\thanks{Ye Han, Lijun Zhang, Dejian Meng, Zhuang Zhang, Xingyu Hu, and Songyu Weng are with School of Automotive Studies, Tongji University, Shanghai, China (e-mail: \{hanye\_leohancnjs, tjedu\_zhanglijun, mengdejian, zhangzhuang, 2410254, 2332944 \}@tongji.edu.cn).}
		
	}
	
	\markboth{Journal of \LaTeX\ Class Files,~Vol.~14, No.~8, August~2024}%
	{Shell \MakeLowercase{\textit{et al.}}: A Sample Article Using IEEEtran.cls for IEEE Journals}


	
	\maketitle
	
	\begin{abstract}
		This paper addresses the problem of long term joint lateral and longitudinal decision-making for multi-vehicle cooperative driving among connected and automated vehicles (CAVs). We propose a Monte Carlo tree search (MCTS) method featuring parallel update mechanism and experiential action preference, with the problem modeled as a finite-horizon, time-discounted multi-agent Markov game. By enabling a single simulation rollout to prune entire families of hazardous actions simultaneously, parallel update mechanism dramatically accelerates the convergence of the search within the vast joint action space. The search is further focused by an experiential action preference heuristic that biases exploration towards efficient maneuvers. Evaluated in randomly generated mixed-traffic scenarios, our method significantly outperforms standard MCTS and reinforcement learning baselines in both safety and traffic efficiency. Crucially, in-depth analysis reveals the discovery of emergent, non-myopic cooperative strategies that are inaccessible to myopic planners. This work provides a computationally tractable solution for complex, multi-agent MCTS, achieving a higher level of strategic cooperative decision-making.
	\end{abstract}
	
	
	\section{Introduction}
\begin{figure}[t]
		\centering
		\includegraphics[width=0.85\linewidth]{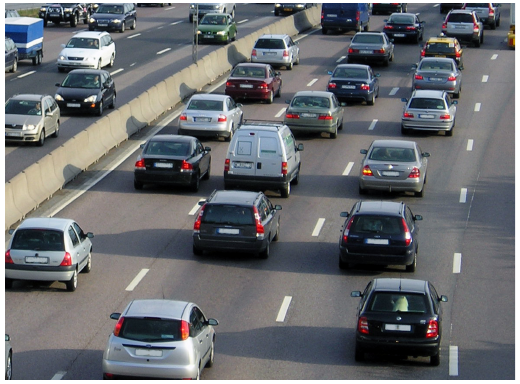}
		\caption{A traffic scenario where cooperation is essential for vehicles to merge safely and efficiently from the rightmost lane. (Image from~\cite{D.T.-42})}\label{fig_real_scene}
	\end{figure}

	\IEEEPARstart{C}{onnected} and Automated Vehicles (CAVs) have emerged as an indispensable component of intelligent transportation systems over the past decade, driven by significant advances in technologies such as vehicle electronics, autonomous driving, Vehicle-to-Everything (V2X) communication, and edge computing~\cite{v2v-1}. For the foreseeable future, autonomous vehicles (AVs) and human-driven vehicles (HDVs) will coexist in urban traffic, creating a complex mixed environment for an extended period~\cite{BabyHomchaudhuri-32},\cite{LeeHess-41},\cite{GuoBan-201}. The self-interested driving behaviors of HDVs and AVs introduce randomness, uncertainty, and even irrationality, posing significant challenges to efficient traffic decision-making~\cite{B.R.-147},\cite{SharmaZheng-167}. Multi-vehicle cooperative decision-making (MVDM) based on V2X communication offers a solution to address the limitations of single-vehicle intelligence, effectively suppressing traffic shockwaves, proactively resolving potential conflicts such as the lane-merging scenario depicted in Fig.~\ref{fig_real_scene}, and enhancing the efficiency, safety, and stability of the entire transportation network \cite{W.M.-40},\cite{M.G.-149},\cite{WangLi-174},\cite{D.T.-42}.
	
	The challenge of multi-vehicle cooperative decision-making has been addressed through several prominent research paradigms. Firstly, optimization-based methods leverage optimal control and mathematical programming to compute ideal vehicle trajectories and control actions~\cite{TypaldosPapageorgiou-165},\cite{ZhangCassandras-161},\cite{BangMalikopoulos-164}. These techniques have been successfully applied to diverse scenarios, including lane-change planning~\cite{TajalliNiroumand-187}, intersection coordination~\cite{R.C.-156}, and on-ramp merging at bottlenecks~\cite{Y.B.-151}. Secondly, learning-based approaches, particularly those centered on deep reinforcement learning (DRL), have gained significant traction~\cite{mv-dl-1}. In this context, agents iteratively refine their decision-making policies through environmental interactions. Multi-agent reinforcement learning (MARL), a natural extension for this domain, has been effectively employed for complex tasks such as trajectory optimization~\cite{FanLu-166},\cite{ S.D.-158}, fleet management~\cite{HanZhang-199}, and integrated traffic control~\cite{ ShiZhou-178}. Finally, game-theoretic frameworks model the interactions among vehicles as a strategic game, wherein each vehicle acts as a rational agent seeking to maximize its individual utility, often with the objective of identifying an equilibrium state~\cite{mixint}\cite{P.C.-215}\cite{HeshamiKattan-173}.
	
	Although these approaches have achieved success, they still exhibit inherent limitations when addressing the complexities of mixed traffic. Optimization-based methods often face a trade-off between solution optimality and computational feasibility, making real-time implementation for non-convex and large-scale problems challenging \cite{opt-1}. DRL methods typically require extensive training, suffer from low sample efficiency, and struggle to provide formal safety guarantees, raising concerns about their reliability in safety-critical scenarios \cite{S.X.-219}. These challenges motivate the exploration of alternative decision-making frameworks. Monte Carlo Tree Search (MCTS), as a heuristic search algorithm, has achieved superhuman performance in complex domains such as Go~\cite{SilverHuang-11}, chess~\cite{SilverHubert-12, SchrittwieserAntonoglou-13}, and real-time strategy games~\cite{VinyalsBabuschkin-49}. The core strengths of MCTS lie in its heuristic nature—requiring minimal prior domain knowledge—and its "anytime termination" property, making it robust to HDV uncertainties and suitable for real-time applications \cite{BrownePowley-10, swiechowskiGodlewski-43}. Through forward-looking simulations, MCTS can reason about the long-term consequences of actions, thereby uncovering complex cooperative strategies that surpass self-interested behaviors~\cite{D.T.-42}.
	
	Building on these notable advantages, Monte Carlo Tree Search (MCTS) represents a promising yet underexplored avenue for MVDM problems. Existing studies have successfully applied MCTS to coordinate vehicles at intersections~\cite{MaYu-184},\cite{H.Y.-217},\cite{MirheliHajibabai-229}, plan lane-changing behaviors~\cite{XuZhang-226}, and organize vehicle platoons~\cite{C.Z.-224}. However, a significant challenge remains: the core difficulty lies in extending MCTS to multi-agent cooperative scenarios, as the joint action space grows exponentially with the number of agents, potentially rendering the search computationally infeasible~\cite{C.Z.-224}. The key open question is how to efficiently explore this vast decision space to discover globally optimal joint strategies rather than locally optimal single-vehicle strategies. Recent review studies highlight that effective MCTS applications in complex real-world domains typically require problem-specific enhancements and hybrid approaches to improve search efficiency~\cite{swiechowskiGodlewski-43}. Therefore, there is a pressing need for MCTS frameworks specifically engineered for multi-agent cooperation to efficiently explore the joint strategy space.
	
	To address this challenge, this paper proposes a parallel update MCTS method tailored for multi-vehicle Markov games in a finite-horizon, discounted setting. The main contributions of this paper are as follows: 
	\begin{itemize}
		\item A novel parallel update MCTS is proposed for multi-vehicle fast joint action space search. The core inovation is a \textit{parallel update} mechanism founded on safety-aware action similarity, which fundamentally increases the information gain of each search iteration compared to existing MCTS algorithms. 
		\item A highly efficient search is achieved by synergistically combining two novel mechanisms. Safety-driven pruning from the parallel update is complemented by efficiency-driven guidance from an \textit{experiential action preference} heuristic. This combination ensures a highly focused and efficient exploration of the vast decision space.
		\item Comprehensive analysis and visualization of the MCTS search process are carried out to explain the proposed algorithm's superior performance. Quantitative metrics and qualitative visualizations provide direct evidence of how the proposed mechanisms achieve a more efficient and focused search.
		\item Complex, emergent cooperative driving strategies are shown to be discovered through rigorous validation. The ability to find these globally optimal solutions, inaccessible to myopic planners, confirms that the proposed method unlock a higher level of strategic multi-vehicle decision making.
	\end{itemize}
	
	The remainder of this paper is organized as follows. Section~\ref{sec_Related} reviews the related work on MVDM and MCTS. Section~\ref{sec_Problem} formulates the MVDM problem and establishes the theoretical framework of the model. Section~\ref{sec_Methodology} details the proposed algorithm, presenting the value-based MCTS method and the parallel update method. Section~\ref{sec_Case} analyzes the algorithm's parameters, evaluates its performance, provides an in-depth interpretation of the model's behavior, and analyzes the resulting cooperative vehicle interactions. Finally, Section~\ref{sec_Conclusion} concludes the paper.
	
	\section{Related Works}\label{sec_Related}
	
	This section systematically reviews the literature on MVDM and MCTS. We begin by surveying the dominant approaches to solving multi-vehicle cooperation problems. Subsequently, we focus on the current state of MCTS applications in the autonomous driving domain, identifying key research gaps to contextualize the motivation and contributions of this paper.
	
	\subsection{Approaches for MVDM in Mixed Traffic}
	
	Research on MVDM can be primarily categorized into optimization-based methods, learning-based methods, and game-theoretic methods.
	
	Optimization-based methods seek theoretically optimal solutions by constructing mathematical models. For example, Zhang and Cassandras~\cite{ZhangCassandras-161} established a decentralized optimal control framework for CAVs at intersections to cooperatively optimize energy consumption and travel time. In the scenario of on-ramp merging, Tang et al.~\cite{TangZhu-230} proposed a hierarchical system optimization model that determines the merging positions and sequence of vehicles by solving a mixed-integer programming problem. Although these methods can guarantee solution optimality, their high computational overhead often renders them impractical for large-scale, non-convex, real-time problems~\cite{opt-1}.
	
	Learning-based methods, particularly deep reinforcement learning (DRL), enable agents to learn driving policies from complex environments~\cite{mv-dl-1},\cite{X.X.-150}. In practice, Fan et al.~\cite{FanLu-166} applied the Multi-Agent Deep Deterministic Policy Gradient (MADDPG) algorithm to enable CAVs to achieve optimal collective utility at mixed-traffic signalized intersections. To suppress disturbances in mixed traffic, Shi et al.~\cite{ShiZhou-178} developed a DRL-based distributed control strategy that allows CAVs to learn and actively counteract traffic oscillations caused by HDVs. However, the opacity of deep learning methods makes it difficult for their policies to provide formal safety guarantees, limiting their trustworthy application in safety-critical scenarios~\cite{S.X.-219}.
	
	Game-theoretic methods provide a powerful analytical framework for strategic interactions among vehicles~\cite{mixint}. For instance, Hang et al.~\cite{P.C.-215} applied coalitional game theory in multi-lane merging zones, allowing CAVs to negotiate safe and efficient driving behaviors by forming coalitions. Similarly, Heshami and Kattan~\cite{HeshamiKattan-173} utilized coalitional games to solve the cooperative lane-changing decision problem for self-organizing CAVs. These methods excel at characterizing interactions among rational agents, but finding and computing stable equilibria can be difficult, and convergence to a global optimum is not guaranteed.
	
	In summary, existing mainstream methods exhibit distinct trade-offs among decision optimality, computational efficiency, and safety guarantees. This creates an opening for the exploration of novel decision-making paradigms that offer a more balanced performance profile and greater adaptability.
	
	\subsection{Application of MCTS in Autonomous Driving}
	
	As a forward-planning heuristic search algorithm, MCTS is well-suited for handling the uncertainty and real-time challenges in autonomous driving due to its minimal reliance on prior knowledge and its "anytime" property~\cite{BrownePowley-10, swiechowskiGodlewski-43}. At its core, MCTS intelligently balances exloration and exploitation using the UCT (Upper Confidence bounds applied to Trees) algorithm~\cite{KocsisSzepesvari-30}, and it has achieved remarkable success in complex domains such as Go~\cite{SilverHuang-11} and real-time strategy games~\cite{VinyalsBabuschkin-49}.
	
	These advantages have motivated numerous researchers to apply MCTS to specific autonomous driving scenarios. For instance, in managing unsignalized intersections, Xu et al.~\cite{H.Y.-217} combined MCTS with heuristic rules to rapidly find a near-optimal vehicle passing sequence within a tree-structured solution space. For the more challenging problem of mandatory lane changes, the same team proposed a bi-level MCTS strategy: the upper level plans for right-of-way, while the lower level resolves lane-changing conflicts, thereby enabling safe cooperative driving~\cite{XuZhang-226}. In the context of vehicle platooning, Liu et al.~\cite{C.Z.-224} designed a decentralized MCTS framework that decomposes the complex multi-vehicle motion planning problem into a series of lane-changing tasks, efficiently organizing scattered CAVs into a platoon.
	
	However, while these studies have achieved encouraging results with MCTS, they typically manage the computational complexity in multi-agent cooperation by simplifying the problem formulation. Specifically, most of these methods rely on decision serialization (i.e., converting a parallel problem into a serial one)~\cite{H.Y.-217, XuZhang-226} or task decomposition~\cite{C.Z.-224}, rather than searching directly in the Joint Action Space of all agents. Although this simplification ensures computational feasibility, it risks overlooking superior cooperative strategies that arise from real-time interactions among agents. Consequently, effectively extending MCTS to directly manage the multi-agent joint action space remains a key bottleneck in the current research landscape~\cite{swiechowskiGodlewski-43}.
	
	To address this challenge, we propose a novel parallel-update MCTS framework. Instead of avoiding the complexity of the Joint Action Space, our approach enhances search efficiency through a specialized parallel update mechanism, designed to uncover higher-level, non-trivial cooperative behaviors among vehicles.
	
	\subsection{Pruning and Parallelization of MCTS}
	
	For context and to provide a key distinction, we briefly introduce two common techniques for enhancing the speed and accuracy of MCTS: pruning and parallelization.
	
	Tree pruning can be divided into two categories: \textit{soft pruning} and \textit{hard pruning}. Soft pruning reduces the risk of optimal actions being prematurely pruned and overlooked. However, some pruning techniques require a reliable state evaluation function, which is not always available in MCTS~\cite{BrownePowley-10,J.Z.-44}. The parallel update method proposed in this paper can be regarded as a more relaxed form of soft pruning, which maximally preserves all potentially promising nodes.
	
	Leaf parallelization, root parallelization, and tree parallelization are the three primary methods for MCTS parallelization~\cite{swiechowskiGodlewski-43}. We mention these techniques here to distinguish them from the parallel update method proposed in this paper. Essentially, these methods increase node visit counts, with each update corresponding to an \textit{actual} node visit. As such, they are orthogonal to and can be combined with the parallel update method presented herein.
	
	\section{Problem Formulation}\label{sec_Problem}
	
	\subsection{Multi-Agent Markov Game}
	
	In this paper, multi-vehicle cooperative driving is modeled as a multi-agent Markov game. A \textit{Markov game} (or \textit{stochastic game})~\cite{kalogiannis2023towards} is specified by a tuple: 
	\begin{equation}
		\langle \mathcal{I},\mathcal{S},\mathcal{A},\mathcal{P},\mathcal{R},\rho_0,\gamma \rangle , 
	\end{equation}
	where $\mathcal{I}$ is the set of agents (in the remainder of this paper, we sometimes use \textit{agent} to refer to CAVs), $\mathcal{S}$ is the state space, $\mathcal{A}$ is the joint action space, $\mathcal{P}$ specifies the state transition probability distribution, $\mathcal{R}$ specifies the reward function, $\rho_0$ denotes the initial state distribution, and $ \gamma \in (0,1]$ denotes a discount factor.
	
	At each time step $t \in \{0,\ldots,T\}$, every agent $i \in \mathcal{I} = \{1,\ldots,N\}$ selects an action according to its state-conditional policy $\pi^{(i)}(a_t^{(i)}|s^{(i)}_t;\bm{\theta}^{(i)})$. Here, $T$ denotes the episode length, $N$ is the number of agents. For agent $i$, $s^{(i)}_t$ is its state at time $t$, and $\bm{\theta}^{(i)}$ are the parameters of its policy. Subsequently, the rewards for all agents, $r^{(1)}_t,\ldots,r^{(N)}_t$, are sampled from the reward distribution $\mathcal{R}(\cdot|s_t,\bm{a}_t)$, and the state transitions to $s_{t+1}$ according to the transition probability function $\mathcal{P}(\cdot|s_t,\bm{a}_t)$. It should be noted that, throughout this paper, we use a parenthesized superscript, as in $x^{(i)}$, to denote a variable $x$ associated specifically with agent $i$, distinguishing it from an exponent.
	
	Although agents receive individual rewards, we are primarily interested in learning \textit{cooperative} behaviors that maximize \textit{total} group return, that is, the sum of all agents' individual rewards across all timesteps. More precisely, we wish to find the optimal agent policy parameters $\bm{\theta}^* = \{\bm{\theta}^{(1)*},\ldots,\bm{\theta}^{(N)*}\} = \underset{\bm{\theta}}{\mathrm{argmax}}  R(\bm{\theta}) $, where
	\begin{equation}
		\label{eq_obj}
		R(\bm{\theta}) = \mathbb{E} \Big[ \sum_{t=0}^T \sum_{j=1}^N \gamma^t r_t^{(j)} \Big| \bm{\pi}_{\bm{\theta}} \Big].
	\end{equation}
	
	This problem formulation is distinct from the typical \textit{greedy} case, where each agent maximizes its own individual return. In this problem formulation, agents should learn to be altruistic in certain situations, by selecting actions that help maximizes group reward, possibly at the expense of some individual reward.
	
	Under the joint strategy $\bm{\pi} = \prod_{i\in \{1,\ldots,N \}}\pi^{(i)}$, we define a $Q$-function with finite time horizon:
	\begin{equation}
		\begin{aligned}
			Q^{\bm{\pi}}(s_t, \bm{a}_t) = &\mathbb{E}_{\bm{\pi}, \mathcal{P}} \left[\frac{ \sum_{k = 0}^{T-t} \gamma^k r_{t+k}(s_{t+k}, \bm{a}_{t+k})} {\sum_{k = 0}^{T-t} \gamma^k} \right]	,
			\label{eq_network_q}
		\end{aligned}
	\end{equation}
	where $r_t = \sum_{j=1}^N r_t^{(j)}$.
	
\begin{figure}[ht]
		\centering
		\includegraphics[width=1.0\linewidth]{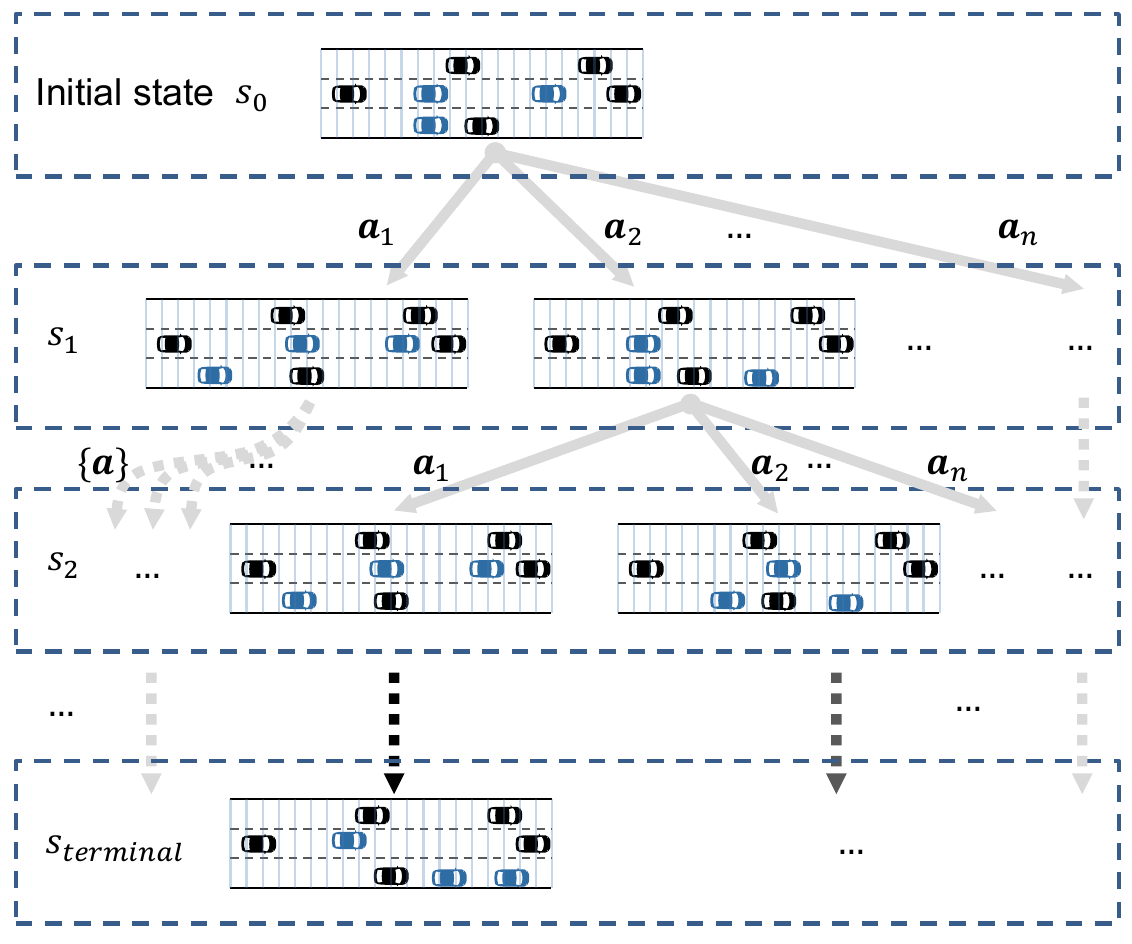}
		\caption{An illustrative partial game tree for the multi-agent Markov game. Each node $\bm{s}_t$ is a system state. Each edge represents a joint action $\bm{a}_t$ taken by all CAVs. For clarity, we only show a limited number of branches and expand the tree from one selected node at each level to illustrate the progression.}\label{fig_game_tree}
	\end{figure}

	In MCTS setting, we use a game tree to represent the extensive-form game (EFG) of MVDM, as illustrated in Fig.~\ref{fig_game_tree}. Each node in the tree represents the system state $s_t$. Under the assumption of a finite and discrete action space, the number of possible system states within a finite time horizon is also finite. The root node $s_0$ represents the initial traffic configuration, while each child node represents a state resulting from taking a joint action $\bm{a}$ from its parent state. The directed edges connecting the nodes represent the joint action $\bm{a}_t$ taken by the CAVs. As defined in this paper, a joint action is a tuple of the individual actions of all CAVs, i.e., $\bm{a}_t = (a_t^{(1)}, a_t^{(2)}, ..., a_t^{(N)})$. As the joint action space is typically vast, the figure provides a simplified and illustrative depiction. 
	For simplicity, Fig.~\ref{fig_game_tree} illustrates deterministic transitions. The figure shows several possible successor states $s_1$ being expanded from $s_0$. To focus on the depth-wise expansion of the tree, we select one of these possible $s_1$ states to expand further, illustrating how executing different joint actions from this state leads to various new states $s_2$. This process continues, with each state node acting as a parent and being expanded downwards, until a predefined terminal state, $s_{\text{terminal}}$, is reached.
	
	In cooperative driving games, we divide all traffic participants into two game coalitions: the CAV coalition and the HDV coalition, which are denoted as $\omega_A$ and $\omega_{-A}$, respectively. The state $s_t$ includes the kinematic information of all vehicles, including both CAVs and HDVs. The actions of HDVs are not part of the joint action $\bm{a}_t$ but are modeled as a stochastic component of the state transition function $\mathcal{P}(s_{t+1}|s_t, \bm{a}_t)$, where $\bm{a}_t$ is the joint action of CAVs only.
	
	This paper employs a value-based Monte Carlo method to derive the driving policy, where at each time step $t$, an action is selected based on a value-based MCTS algorithm, which is summarized in Algorithm~\ref{alg_vb_mcts}. The $\textsc{SelectNode}$ function typically selects the node with the highest UCB value for tree traversal, while the final action is chosen based on the maximum $Q$-value. The $\textsc{ExpandNode}$ and $\textsc{BackPropagation}$ procedures will be explained in detail in subsequent sections.
	
	\begin{algorithm}[H]
		\caption{Value-Based MCTS Algorithm.}\label{alg_vb_mcts}
		\begin{algorithmic}[1]
			\STATE {\textsc{GetAction}}$(s)$
			\STATE \hspace{0.5cm} creat root node $\mathcal{N}_0\{\bm{a}_0, Q_0, n_0\}$ with traffic state $s$
			\STATE \hspace{0.5cm} \textbf{while} within maximum rollout \textbf{do}
			\STATE \hspace{1.0cm} $\mathcal{N}\{\bm{a}, Q, n\}$ $\gets$ $\mathcal{N}_0\{\bm{a}_0, Q_0, n_0\}$
			\STATE \hspace{1.0cm} \textbf{while} $\mathcal{N}\{\bm{a}, Q, n\}$ is nonterminal \textbf{do}
			\STATE \hspace{1.5cm} \textbf{if} $\mathcal{N}\{\bm{a}, Q, n\}$ has no childen nodes \textbf{then}
			\STATE \hspace{2.0cm} $\textsc{ExpandNode}(\mathcal{N}\{\bm{a}, Q, n\})$
			\STATE \hspace{2.0cm} \textbf{break}
			\STATE \hspace{1.5cm} \textbf{else} $N$ $\gets$ $\textsc{SelectNode}(\mathcal{N}\{\bm{a}, Q, n\})$
			\STATE \hspace{1.5cm} \textbf{end if}
			\STATE \hspace{1.0cm} \textbf{end while}
			\STATE \hspace{1.0cm} $\textsc{BackPropagation}(\mathcal{N}\{\bm{a}, Q, n\})$
			\STATE \hspace{0.5cm} \textbf{end while}
			\STATE \hspace{0.5cm} \textbf{return} $\textsc{SelectNodeByQ}(\mathcal{N}_0)$
		\end{algorithmic}
	\end{algorithm}
	
	\subsection{World Model}\label{sec_world_model}
	
\begin{figure}[ht]
		\centering
		\includegraphics[width=1.0\linewidth]{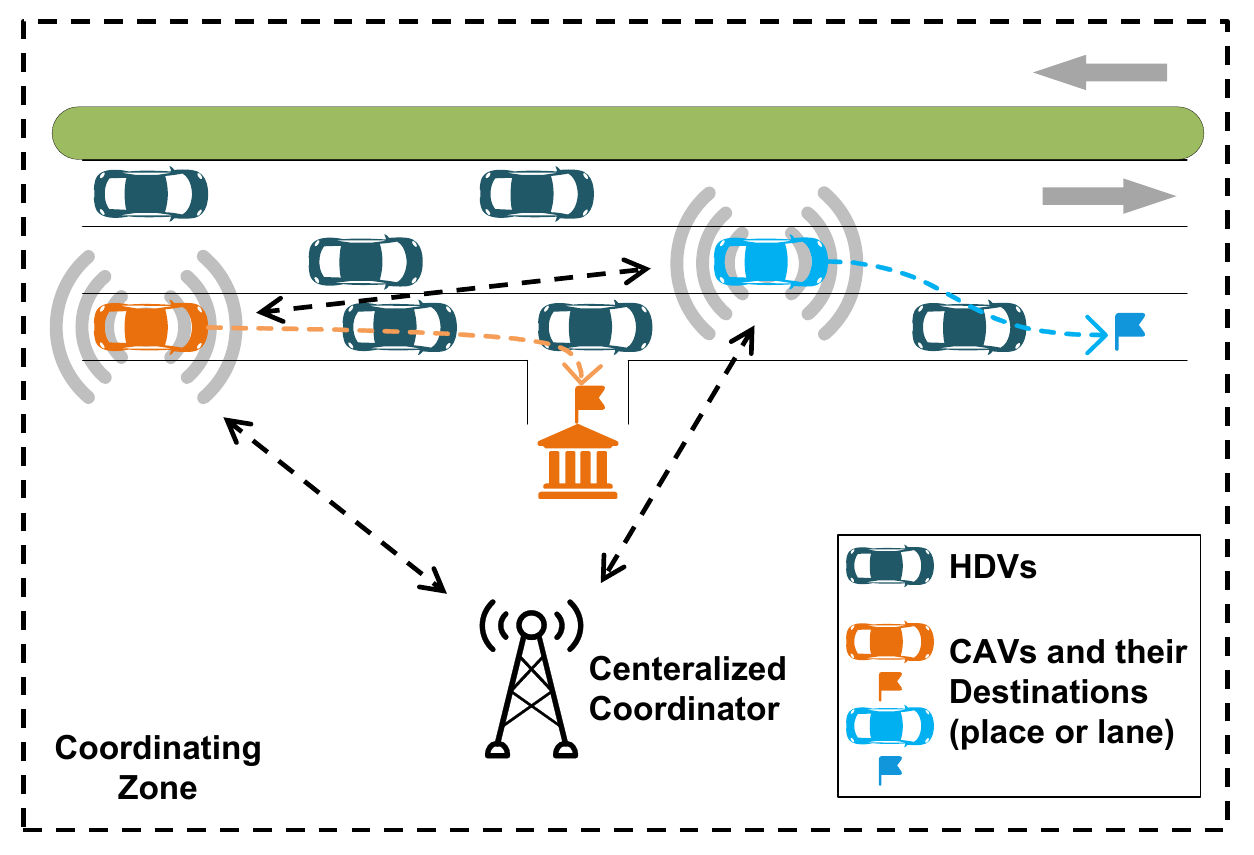}
		\caption{Basic scene of multi-vehicle cooperative driving. In the coordinating zone, a centralized coordinator collects real-time information and sends optimal driving strategies to CAVs. The figure showcases different driving intentions for CAVs, such as exiting the road or changing to a target lane}\label{fig_basic_scene}
	\end{figure}

	The basic scenario of multi-vehicle cooperative driving is shown in the Fig.~\ref{fig_basic_scene}. Consider a one-way, three-lane urban road with a length of 300 meters. The coordinating zone in the figure represents the decision-making area of interest, where all CAVs are controlled by the centeralized coordinator. The CAVs in this area have different driving intentions, which could be a target position (e.g., exiting the road at a specific location, as illustrated by the orange vehicle in Fig.~\ref{fig_basic_scene}.), or a target lane (serving as the entry position for the vehicle into the next coordinating zone).
	
	In this article, we make the following assumptions: 
	\begin{enumerate}
		\item The state of the vehicle, including position, speed, and destination (or intention), can be received in real time by the centeralized coordinator. The coordinator gives the optimal driving strategy of the CAVs based on the information, and the CAVs always accepts the instruction and execute it;
		\item The HDVs in the traffic environment are homogeneous. The models described in the following section can represent human driving behaviors. 
	\end{enumerate} 
	
	This paper employs a rule-based microscopic traffic model to describe the behavior of HDVs, thereby providing a dynamic and reproducible testbed for the CAV's decision-making algorithm. For longitudinal car-following behavior, we adopt the improved Krauss model~\cite{SongWu-27}. Krauss model~\cite{etde_627062} are widely used in traffic flow simulation due to its collision-free property and computational simplicity. While its default parameters may not fully replicate all the nuances of real-world driving, the model effectively simulates the fundamental reactive behaviors of human drivers under safety distance and desired velocity constraints. This capability is crucial for MCTS algorithms, which depend on extensive forward simulations. The improved Krauss model enhances traffic flow realism and behavioral diversity through a driver imperfection mechanism while preserving the collision-free property and computational efficiency of the original model.
	
	In Krauss model, for every time step, the desired velocity, 
	\begin{equation}
		\tilde{v}(t + \Delta t)=\min \left[v_{\max }, v+a\Delta t, v_{\text {safe }}\right], 
	\end{equation}
	where 
	\begin{equation}
		\begin{aligned}
			v_{\text {safe }}=-b \tau _k +\sqrt{(b\tau_k)^{2}+v_l^2 + 2bg} ,
		\end{aligned}
	\end{equation}
	$b$ is the maximum deceleration, $\tau_k$ is the reaction time, and $v_l$ is the speed of the leading vehicle. The driver imperfection mechanism, acknowledging that drivers do not always execute optimal decisions with perfect precision, is achieved by introducing a stochastic noise term into the velocity update ~\cite{SongWu-27}. The update rule can be expressed as:
	\begin{equation}
		v_{\text{des}} = \widetilde{v} - a_{\max}\Delta t \cdot \varepsilon \cdot \xi
	\end{equation}
	where $\xi$ is a random number drawn from the interval $[0, 1]$, and $\varepsilon$ is a parameter controlling the magnitude of the noise. This improvement provides a solid testbed for evaluating the robustness and effectiveness of our CAV cooperative decision-making algorithm.
	
	The lane changing model of HDVs adopts LC2013 lane changing model~\cite{Erdmann-28}. The hierarchical architecture lane changing model considers 4 steps at each moment: 
	\begin{enumerate}
		\item Calculate the best follow-up lane;
		\item Suppose that the current lane is maintained and the safe speed is calculated by combining the lane change correlation speed at the previous moment;
		\item Calculating lane change requests;
		\item Execute the director action or calculate the speed requirement at the next moment (may plan the speed for a period of time in advance). Whether to change the speed depends on the urgency of the lane change.
	\end{enumerate}
	
	By integrating the improved Krauss model with the LC2013 lane-changing model, we can construct a high-fidelity interaction environment with HDVs for the CAV agents.
	
	\subsection{Action Space}
	
	For each vehicle, we consider both the lateral and longitudinal behaviors. Longitudinal actions include accelerating (AC), speed keeping (SK) and decelerating (DC), and lateral actions consist of left lane changing (LC), lane keeping (LK) and right lane changing (RC). Considering that longitudinal and lateral actions can be performed at the same time, there are 9 action for a single vehicle, 
	\begin{equation}
		\mathcal{A}^{(i)}=\left\{\left(a_{\text {lon}}, a_{\text {lat}}\right) | a_{\text {lon}} \in A_{\text {lon}}, a_{\text {lat}} \in A_{\text {lat}}\right\},
	\end{equation}
	where $A_{\text{lon}}=\{\text{AC}, \text{SK}, \text{DC}\}$, and $A_{\text{lat}}=\{\text{LC}, \text{LK}, \text{RC}\}$. Then the joint action space,
	\begin{equation}
		\mathcal{A} = \prod_{i\in \{1,\ldots,N \}}\mathcal{A}^{(i)}.
	\end{equation}
	
	\subsection{Reward Function}
	
	Our work aims at the driving efficiency and safety of global traffic in concerned area. The reward function is designed following Equation~(\ref{eq_rewardfun}),
	\begin{equation}
		\begin{aligned}
			r&=w_{1} R_{\text{speed}}+w_{2} R_{\text {intention }}+w_{3} P_{\text {collision }}+w_{4} P_{LC} \\
			&=\frac{1}{N}(w_{1} {\sum_{i=1}^{N} r_{\text{speed}}^{(i)}}+w_{2} N_{\text {arrived }}+w_{3} N_{\text {collision }}+w_{4} N_{L C})
		\end{aligned}
		\label{eq_rewardfun}
	\end{equation}
	where $N$ is the number of vehicles (including HDVs and CAVs), $N_{\text {arrived }}$ is the vehicle passing through intention area at the previous time step and aiming for the ramp, $N_{\text {collision }}$ is the number of vehicles involved in collision, and $N_{L C}$ is the number of frequently lane-changing vehicles.
	
	In the previous work we have reviewed, $R_{\text {speed }}$ are typically represented directly by the vehicles' speed or a linear combination of it. However, in our work, due to the simulation's small time step and the vehicles' inherent baseline speed, this reward design approach often fails to adequately capture the differences between actions (we refer to this characteristic as  \textit{partial-steady-state}). In this paper, we propose a reward function design tailored for partial-steady-state systems. To improve traffic efficiency, we assign a reward value to speed increasing. When a vehicle's speed reaches a threshold, we assign the same reward to speed-keeping actions, that is,
	\begin{equation}
		r_{\text{speed}}^{(i)}=\left\{
		\begin{array}{l}
			\begin{aligned}
				& r_{\text{speed}} &,&\text{ if $a^{(i)} > 0$ or }a^{(i)} = 0\text{ and }v>v_{\text{thres}}\\
				& 0 &,&\text{ otherwise}
			\end{aligned}
		\end{array}\right.
	\end{equation}
	
	For intention reward and collision penalty, it is consistent with the settings of \cite{rewardfun}. In this paper, we do not punish frequent lane change, and instead give a small reward to lane keeping.
	
	\begin{figure*}[htbp]
		\centering
		\includegraphics[width=0.8\linewidth]{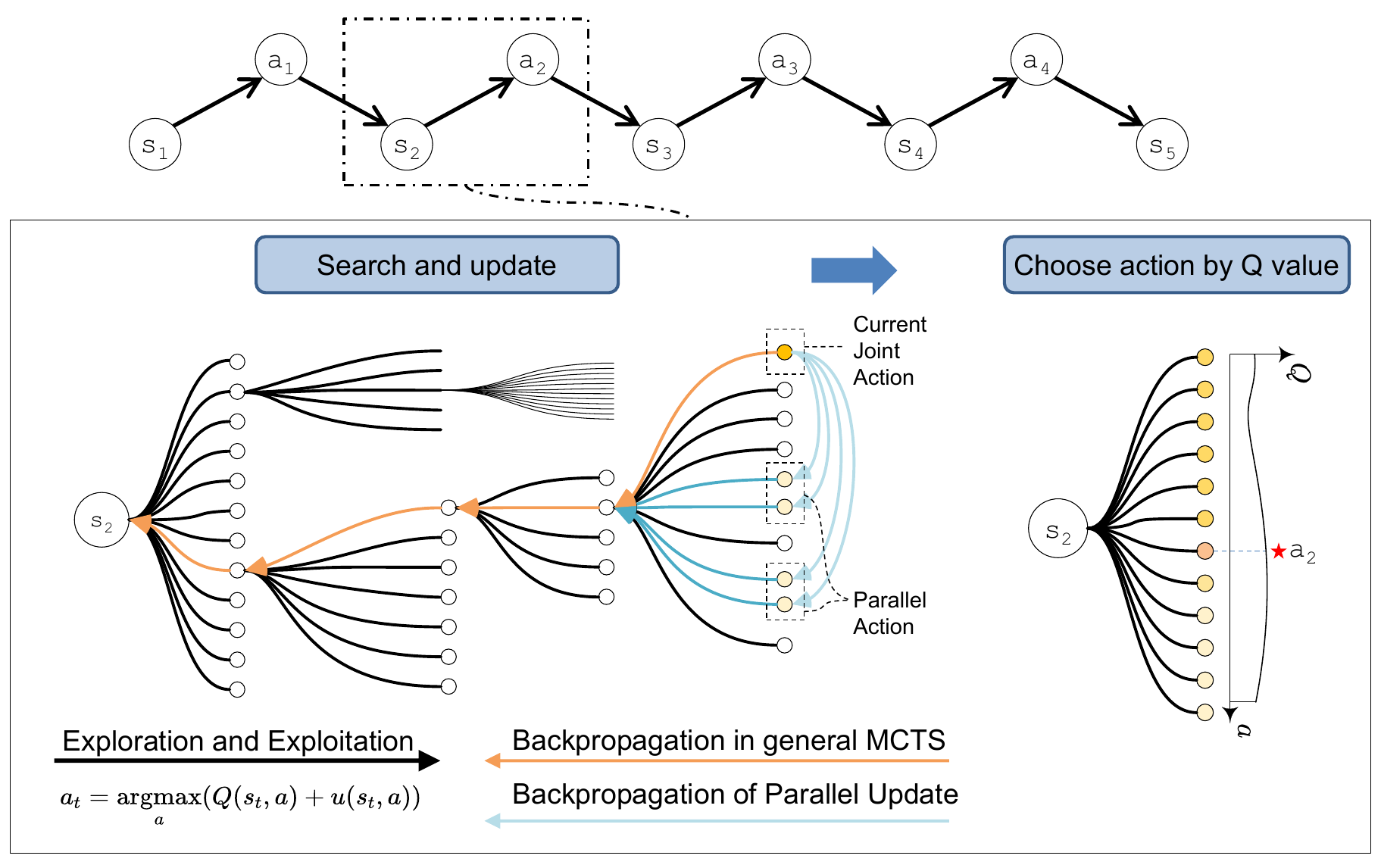}
		\caption{
			The proposed Parallel Update MCTS framework. The algorithm iteratively performs search at each decision step (e.g., $s_2$). During the backpropagation stage, our key contribution, the Parallel Update mechanism (light blue arrows), enables a single simulation to update multiple branches of the search tree, a significant departure from the standard single-path update (orange arrow). This accelerates convergence in the vast joint action space. The best action is selected based on the resulting Q-values.
		}\label{fig_tree_search_detail}
	\end{figure*}
	
	\section{Methodology}\label{sec_Methodology}
	
	\subsection{MCTS Method}
	
	MCTS generally consists of four steps: selection, expansion, simulation and back propagation. 
	\begin{enumerate}
		\item{\textbf{Selection}: Start from the root node, select the child nodes continuously (Algorithm~\ref{alg_vb_mcts}, \textsc{SelectNode}), until reaching a leaf node, and then update the tree depending on that node. A typical rule is
			\begin{equation}
				\bm{a}_{t}=\underset{\bm{a}}{\operatorname{argmax}}\left(Q\left(s_{t}, \bm{a}\right)+u\left(s_{t}, \bm{a}\right)\right),
			\end{equation}
			where 
			\begin{equation}
				u(s, \bm{a}) \propto \sqrt{\frac{\log{n_p}}{(1+n)}}. 
			\end{equation}
			A typical form of $u$ is $c(s) p(s,\bm{a}) \sqrt{\frac{\log{n_p}}{(1+n)}}$. Here, $c(s)$ is a coefficient and $p(s,\bm{a})$ is the prior probability of action selection. The root node is the current game state, and the leaf node is any potential child node that has not yet been explored.}
		\item{\textbf{Expansion}: Generate new child nodes. May initially exclude some illegal of obviously not resonable actions. }
		\item{\textbf{Rollout}: A random simulation, simulate a game from the current state until terminal state. It should be noted that if the current node is new, expand is performed, and if the node has been updated, rollout is performed.}
		\item{\textbf{Back Propagation}: Use the value of leaf node $Q$ to update all nodes on the path to the root node.}
	\end{enumerate}
	
	\subsection{Value Based Settings}
	
	The value-based method comes from reinforcement learning. In value function of RL, the agent adopts a strategy from the value function, which is an indirect method to generate the strategy. The true value of an action is the expected reward when the action is selected. 
	MCTS provides a way to estimate this value by averaging the outcomes of many simulated trajectories (rollouts). A straightforward sample-average estimate for the value of taking action $\bm{a}$ from a state $s$ is
	\begin{equation}
		\hat{Q}(s, \bm{a}) = \frac{1}{n_\text{rollout}(s,\bm{a})} \sum_{k=1}^{n_\text{rollout}(s,\bm{a})} G_k,
		\label{eq_sample_average_q}
	\end{equation}
	where $n_\text{rollout}(s,\bm{a})$ is the number of times action $\bm{a}$ has been tried from state $s$, and $G_k$ is the $k$-th caculation of Equation~(\ref{eq_network_q}) based on simulation that started from $(s, \bm{a})$.
	With number of rollouts increasing, $ \hat{Q}$ converges to $Q$. For all rollouts, we introduce a rollout depth indicator $i$ and its maximum value $i_{\text{max}}$, the set of nodes visited at depth $i$ is $\bm{D}_{i}$, Equation~(\ref{eq_sample_average_q}) could be written as
	\begin{equation}
		\begin{aligned}
			\hat{Q}(s,\bm{a}) = \frac{\sum_{i=0}^{i_{\text{max}}}\left( \gamma^{i}\sum_{j \in \bm{D}_{i}} r_{j}\left(\bm{a}_{j}\right)\right)}{\sum_{i=0}^{i_{\text{max}}} \gamma^{i}\left|\bm{D}_{i}\right|}
		\end{aligned}
		\label{eq_def_Q_discount}
	\end{equation}
	
	\subsection{Parallel Update in Value Based MCTS}
	
	For decision-making problems with large action spaces, search tree pruning has been a long-standing focus of research. Appropriate model pruning methods can significantly improve search speed without substantially affecting decision quality (or obtain strategies of the same quality with lower search costs). The parallel update method proposed in this paper can be seen as a softened version of soft pruning, which maximizes the retention of exploration possibilities for all nodes.
	
	We start introducing the parallel update method by defining \textit{parallel actions}. Parallel actions refer to those that are similar from a safety perspective. 
	We use the representative Time-to-Collision (TTC) to quantify the similarity of agent actions. TTC is one of the key quantitative indicators of vehicle safety. It represents the time required for a vehicle to collide with an obstacle or another vehicle ahead, assuming constant speed and direction.
	\begin{equation}
		\mathrm{TTC}=\frac{D}{V} ,
	\end{equation}
	where $D$ is the distance between two vehicles or vehicles and obstacles. $V$ is the relative speed. 
	
	For the two actions of acceleration and speed maintenance, the relative error of TTC between can be caculated as
	\begin{equation}
		\begin{aligned}
			\operatorname{err}(\tau,V) & =\frac{\frac{S-V \cdot \tau}{V}-\frac{S-\left(V+\frac{1}{2} a \cdot \tau\right) \tau}{V+a \cdot \tau}}{\frac{S-V \cdot \tau}{V}} \\
			& = \frac{1 + \frac{1}{2} \cdot \frac{V\tau}{S - V\tau}}{1 + \iota}
		\end{aligned}
		\label{eq_action_simi}
	\end{equation}
	where $S$ is the current distance between vehicles, and $\tau$ denotes the decision interval, $V$ is the relative speed, $V = v_s - v_p$, and $a$ is the relative acceleration. $\iota = V / a \tau$.
	
	\begin{figure}[h]
		\centering
		\includegraphics[width=0.9\linewidth]{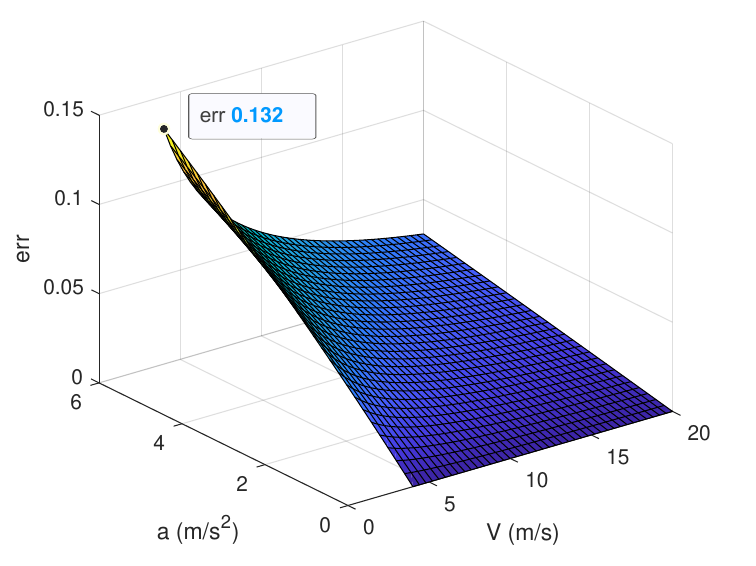}
		\caption{Relative error of Time-to-Collision (TTC) between 'accelerate' and 'maintain speed' actions. The plot is generated using Equation (\ref{eq_action_simi}) with an initial separation of $S = 20$ m and a decision time step of $\tau = 0.1$ s.}\label{fig_action_simi}
	\end{figure}
	
	We caculated the relative TTC error between two actions according to Equation~(\ref{eq_action_simi}), and the result is shown in Fig.~\ref{fig_action_simi}. The update frequency for autonomous vehicle path planning and decision-making typically ranges from 10 to 50 Hz~\cite{Padencap-55}, corresponding to a time step range of 0.02 to 0.1 seconds. For this plot, we assume a typical scenario with an initial separation $S = 20$ m and a decision interval $\tau = 0.1$ s. The normal driving speed of vehicles on urban arterial roads ranges from 30 to 60 km/h (8.33 to 16.67 m/s). To ensure passenger comfort during smooth starts and acceleration on urban roads or highways, the acceleration is typically maintained within the range of 1.0 to 3.5 m/s\textsuperscript{2}. Within these parameter ranges, it is evident that the maximum TTC error is less than 15\%. We can therefore consider the two actions to be \textit{similar} in terms of safety.
	
	In fact, this similarity is more pronounced in trial-and-error-based algorithms without prior knowledge. For example, when a vehicle does not have sufficient conditions for lane changing, the effects of different actions during the lane change process, whether it is lane-change-acceleration, lane-change-deceleration, or lane-change-maintain, are generally \textit{similar} (all of which could potentially lead the vehicle into a hazardous state, especially in conservative driving styles). Additionally, the effects of acceleration and deceleration become more pronounced over the long-term subsequent steps, whereas their impact at the current time step is minimal.
	
	Based on the above analysis, we give the definition of parallel actions in joint action space:
	\begin{definition}[Parallel Action]\label{def_simi_action}
		In the joint action space $\mathcal{A}=\left(\mathcal{A}^{(1)}, \mathcal{A} ^{(2)},\ldots, \mathcal{A}^{(N)}\right)$, for the joint action $\bm{a}_i=\left({a}^{(1)}_i, {a} ^{(2)}_i,\ldots, {a}^{(N)}_i\right)$, if there exists $a_j^{(k)} \sim a_i^{(k)}$ ($\sim$ denotes two similar actions, $j \neq i$, $k \in \{1,2,\ldots,N\}$), and vehicle $i$ received negative feedback for performing $a_i^{(k)}$, then the joint action $\bm{a}_j$ is \textit{parallel action} of $\bm{a}_i$, that is, $ \bm{a}_j \sim \bm { a }_i $.
	\end{definition}
	
	With Definition \ref{def_simi_action}, the MCTS with parallel update method can be represented by  Algorithm~\ref{alg_vb_mcts_para} in each step of decision-making.
	
	\begin{algorithm}[h]
		\caption{Back propagation with parallel update.}\label{alg_vb_mcts_para}
		\begin{algorithmic}[1]
			\STATE {\textsc{BackProbagation}}$(N\{\bm{a}, Q, n\})$
			\STATE \hspace{0.5cm} \textsc{Update}$(N\{\bm{a}, Q, n\}, Q)$
			\STATE \hspace{0.5cm} \textbf{for} $N_p\{\bm{a}_p, Q_p, n_p\}$ in BrotherNodes of $N\{\bm{a}, Q, n\}$
			\STATE \hspace{1.0cm} \textbf{if} $\bm{a}_p \sim \bm{a}$ \textbf{then}
			\STATE \hspace{1.5cm} \textsc{Update}$(N_p\{\bm{a}_p, Q_p, n_p\}, Q, \text{Parallel=True})$
			\STATE \hspace{1.0cm} \textbf{end if}
			\STATE \hspace{0.5cm} \textbf{end for}
		\end{algorithmic}
	\end{algorithm}
	
	The function \textsc{Update} in Algorithm \ref{alg_vb_mcts_para} is designed to be recursive, as shown in Algorithm \ref{alg_vb_node_update}.
	
	\begin{algorithm}[h]
		\caption{Node update.}\label{alg_vb_node_update}
		\begin{algorithmic}[1]
			\STATE {\textsc{Update}}$(N\{\bm{a}, Q, n\}, Q_l, \text{Parallel=False}, \text{Layer=0})$
			\STATE \hspace{0.5cm} \textbf{if} \text{Parallel} \textbf{then}
			\STATE \hspace{1.0cm} $\text{eff} = \gamma _{p} \cdot \gamma ^{\text{Layer}}$
			\STATE \hspace{0.5cm} \textbf{else}
			\STATE \hspace{1.0cm} $n = n + 1$
			\STATE \hspace{1.0cm} $\text{eff} = \gamma ^{\text{Layer}}$
			\STATE \hspace{0.5cm} \textbf{end if}
			\STATE \hspace{0.5cm} $\text{Re}_N = \text{Re}_N + \text{eff}$
			\STATE \hspace{0.5cm} $\text{Rt}_N = \text{Rt}_N + \text{eff} \cdot Q_l$
			\STATE \hspace{0.5cm} $Q = {\text{Rt}_N}/{\text{Re}_N}$
			\STATE \hspace{0.5cm} \textbf{if} $N\{\bm{a}, Q, n\}$ is not root node \textbf{then}
			\STATE \hspace{1.0cm} $\text{Layer} = \text{Layer} + 1$
			\STATE \hspace{1.0cm} \textsc{Update}(\textsc{Parent}($N\{\bm{a}, Q, n\}$), $Q_l$, $\text{Layer}$)
			\STATE \hspace{0.5cm} \textbf{end if}
		\end{algorithmic}
	\end{algorithm}
	
	As a result, the $Q$ value defined in Equation~(\ref{eq_def_Q_discount}) is evaluated as
	\begin{equation}
		\begin{aligned}
			\hat{Q}(\bm{a})= \frac{\sum_{i} \left(\gamma^{i}\sum_{j \in \bm{D}_{i}} \left( r_{j}\left(\bm{a}_{j}\right)+\gamma_{p} \sum_{\bm{P}_{j}} r_{j}\left(\bm{a}_{j, p}\right)\right)\right)}{\sum_{i} \gamma^{i}\left(\left|\bm{D}_{i}\right|+\gamma_{p}\sum_{j \in \bm{D}_{i}}\left|\bm{P}_{j}\right|\right)}
		\end{aligned}
		\label{eq_para_update}
	\end{equation}
	where, $i$ is the rollout depth indicator, $\bm{D}_{i}$ represents the set of nodes directly visited at step $i$ of the rollout process, and $\bm{P}_{j}$ represents the set of parallel nodes of nodes $j$ in $\bm{D}_{i}$ accessed at depth $i$. $\bm{a}_{i}$ and $\bm{a}_{j,p}$ denote the actions corresponding to these nodes. As seen from Equation~(\ref{eq_para_update}), the $Q$-value of a node includes information from nodes directly explored during the rollout process as well as information from parallel actions. Compared with the algorithm without parallel update, the number of $Q$ value update is increased by $\sum_{i} \sum_{j \in \bm{D}_{i}} |\bm{P}_{j}|$, which helps boosting dangerous action exclusion.
	
	In Algorithm \ref{alg_vb_node_update}, it should be noted that parallel update will only update its sibling nodes synchronously when directly accessing the dangerous nodes, and will not be triggered in other cases.
	
	\begin{figure*}[htbp]
		\centering
		\includegraphics[width=0.8\linewidth]{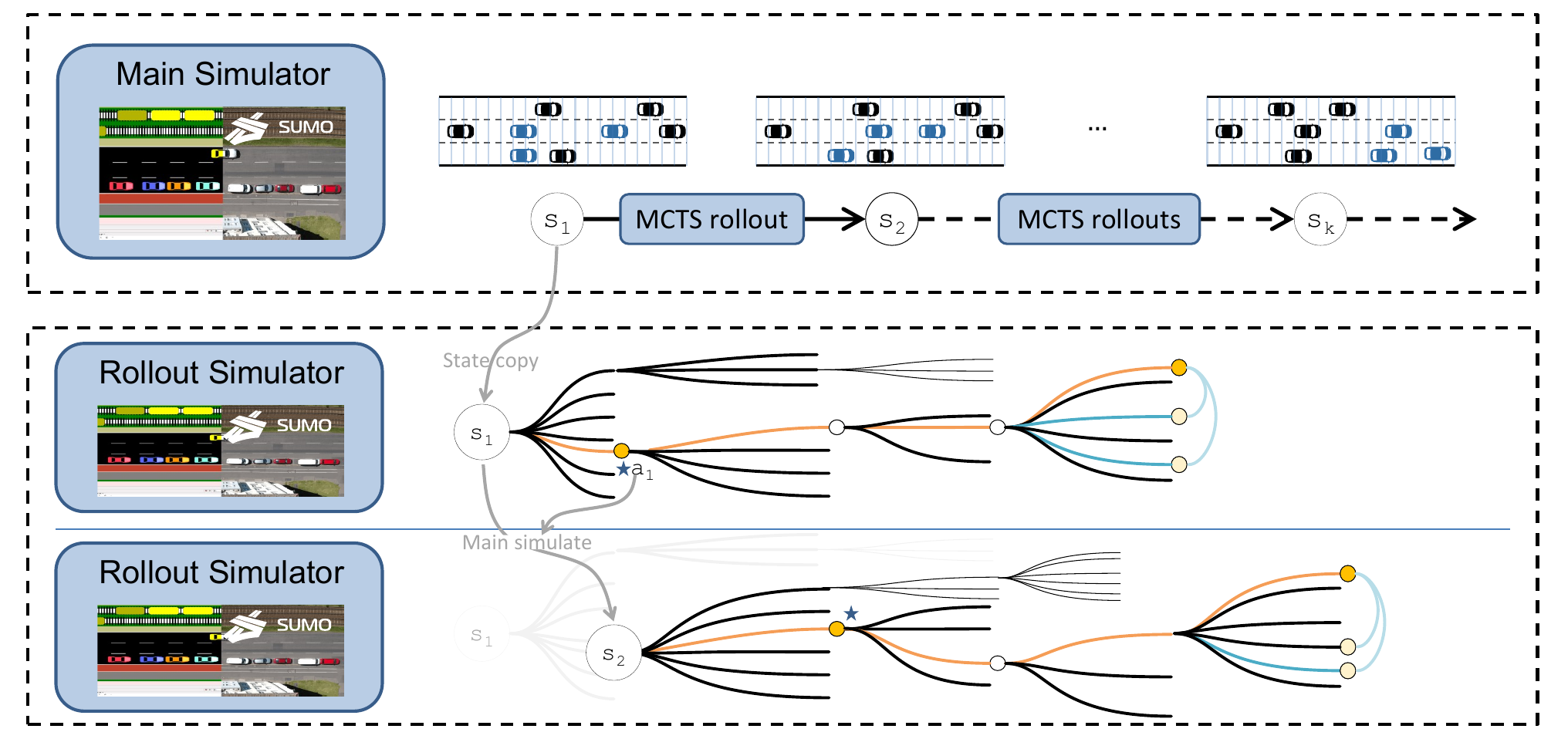}
		\caption{
			Architecture of the bi-layer MCTS co-simulation loop. Simulations are carried out in SUMO based framework. The diagram illustrates the separation between the main simulation and the MCTS rollout process. At each time step $t$, The current environment state ($s_t$) is duplicated into a temporary SUMO simulation instance. The MCTS Controller uses this sandbox to perform numerous rollouts, building a search tree to evaluate future actions without affecting the main simulation.
		}\label{fig_bi_layer_sim}
	\end{figure*}
	
	\section{Case Study}\label{sec_Case}
	
	\subsection{Case Settings}\label{sec_case_setting}
	
	We employ the Flow framework~\cite{WuKreidieh-57} to construct our simulation scenarios and validate the proposed algorithm. Flow is a computational framework designed for experiments in DRL and traffic control, providing a suite of benchmark scenarios and tools for designing custom traffic environments. The underlying microscopic simulation is powered by SUMO (Simulation of Urban MObility), a widely adopted and robust traffic simulator. The HDV models within the simulation are configured as described in Section~\ref{sec_world_model}. To rigorously test the planning capabilities of the proposed algorithm, we disable all of SUMO's default collision avoidance mechanisms and automatic safety constraints for the CAVs, ensuring that any observed safe driving behaviors are solely the result of the MCTS planner's decisions, rather than the simulator's built-in safety features.
	
	At the core of our validation platform is a bi-layer co-simulation loop with a time step of 0.1s, as illustrated in Fig.~\ref{fig_bi_layer_sim}. The process at each time step strictly separates the advancement of the main simulation from the MCTS search procedure:
	\begin{enumerate}
		\item \textbf{State Duplication}: At the beginning of a new time step in the main simulation, the complete traffic state (including the precise position, speed, and lane ID of all vehicles) is duplicated into a temporary independent SUMO instance, which serves as a sandbox for the MCTS algorithm's exploration.
		\item \textbf{MCTS Search}: Within the duplicated instance, the MCTS algorithm executes its rollouts. Each rollout is a fast forward simulation of a potential future joint action sequence. This trial-and-error process allows the algorithm to evaluate the long-term value of different decisions without affecting the main simulation's state.
		\item \textbf{Action Selection and Execution}: Once the search is complete, the MCTS controller selects the joint action with the highest Q-value. This chosen action is then passed to the main simulator for execution. SUMO subsequently advances the main simulation by a single time step.
		\item \textbf{Incremental Tree Update}: The MCTS search tree is persistent. After an action is executed, the root of the tree is advanced to the child node corresponding to the executed action. This preserves the entire subtree of prior search information, providing a warm start for the next decision cycle.
	\end{enumerate}
	This closed-loop interaction continues until the simulation episode concludes.
	
	\begin{table}[h]
		\centering
		\caption{Parameters settings for simulation}\label{agentparameter}
		\begin{tabular}{lc}
			\toprule
			\multicolumn{1}{c}{Parameters} & \multicolumn{1}{c}{Value} \\ \midrule
			Number of HDV and CAV       & 4, 2                        \\
			Departure speed of HDV and CAV  & 10 m/s, 10 m/s                      \\
			Acceleration value of CAV        & 3.5 m/s\textsuperscript{2}                   \\
			Max speed of HDV and CAV       & 30 m/s, 30 m/s                     \\
			Simulation interval $\tau$     & 0.1 s \\
			$b$, $t_k$ $\varepsilon$ & 9 m/s\textsuperscript{2}, 1.1 s, 0.5 \\
			$r_{\text{speed}}$ & 10 \\
			$v_{\text{thres}}$ & 28 m/s \\
			$\gamma$, $\gamma_{p}$&0.99, 0.01\\
			$w_{1}$, $w_{2}$, $w_{3}$, $w_{4}$ & 1, 30, -50, 2 \\
			\bottomrule
		\end{tabular}
	\end{table}
	
	The simulation environment consists of a 300-meter, three-lane, unidirectional road segment, as illustrated in Fig.~\ref{fig_basic_scene}. The objective for CAV 1 is to reach a target point at the 150-meter mark, while CAV 2 is tasked with merging into the rightmost lane to exit the roadway. Detailed vehicle dynamics parameters, reward function weights, and simulation settings are provided in Table~\ref{agentparameter}. The initial positions of the vehicles are generated using a fixed random seed of 42, and a total of 200 experimental runs are conducted for each evaluated method. The weights for the reward function were determined based on a principled hierarchy of driving objectives rather than an exhaustive search, ensuring that the agent's behavior aligns with logical driving priorities. Foremost among these is safety related $w_3$, a strong negative signal guarantees that when a single hazardous action is penalized, the negative reward propagated to its parallel actions is significant enough to rapidly prune large, unsafe regions of the joint action space, which is a core tenet of our algorithm's efficiency. Task completion is the second most critical factor. The intention reward $w_2$ is set to 30, providing a strong positive incentive for the agents to achieve their primary missions. Finally, the third-tier priorities of traffic efficiency and behavioral smoothness are encoded with smaller weights. ($r_{\text{speed}}=10$, $w_1=1$, $w_4=2$) We provide further analysis of parameter impacts in Section~\ref{sec_hyper_sens}.
	
	\subsection{MCTS Implementation Details}
	
	\textbf{Node Properties}: In this paper, we assign 7 variables to a node, which could be denoted by a tuple:
	\begin{equation}
		N\{\bm{a}, Q, n\} := \langle P, p, C, n, u, \mathrm{Re}, \mathrm{Rt}, Q\rangle
		\label{eq_node_attr}
	\end{equation}
	where $P$, $p$, $C$, $n$, $u$, $\mathrm{Re}$, $\mathrm{Rt}$, $Q$ are the node's parent node, prior probablity, first-level child nodes, times of direct access, $\mathrm{UCB}$ value, sum of coefficient of rollout returns, rollout return value, and $Q$ value, respectively.
	
\begin{figure}[h]
		\centering
		\includegraphics[width=0.7\linewidth]{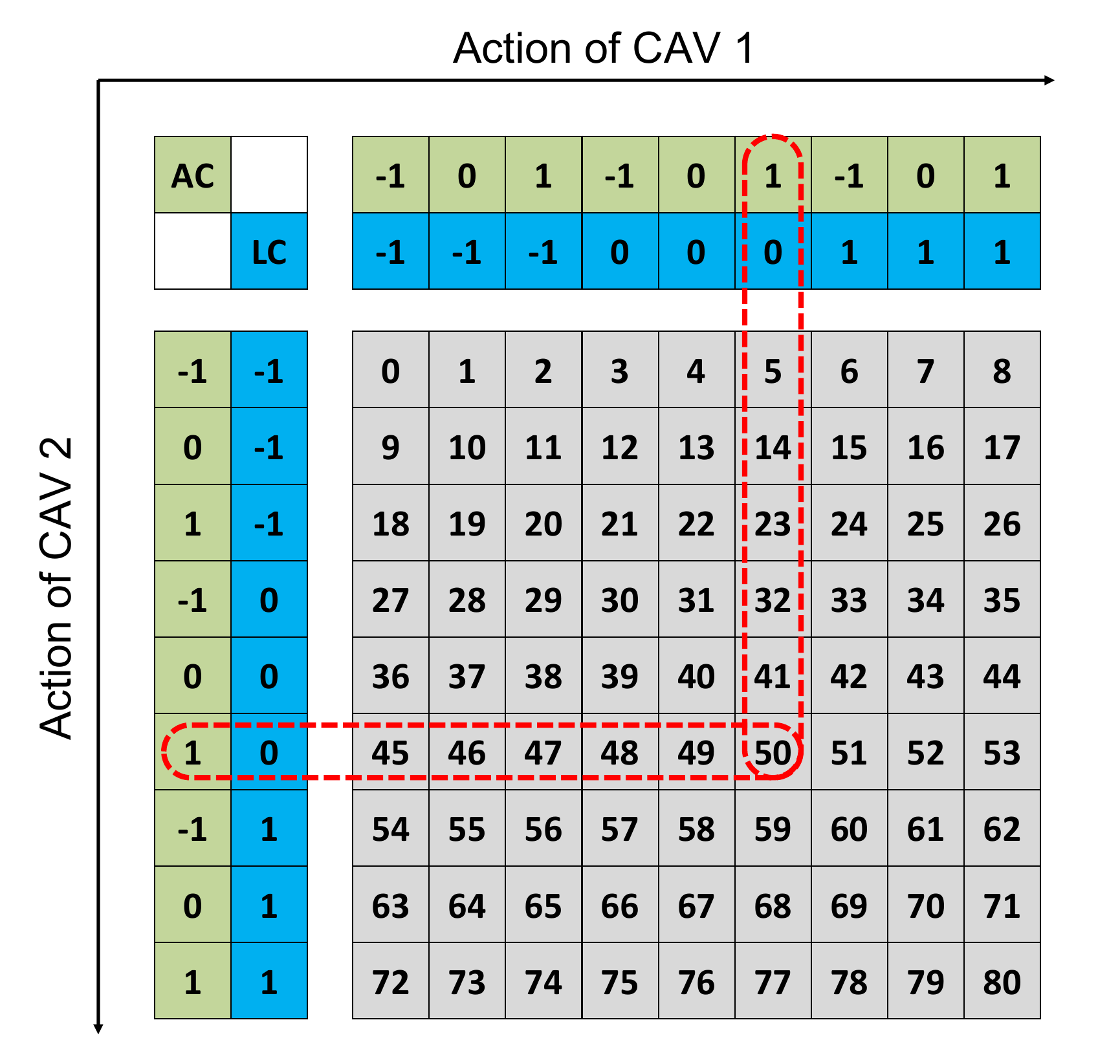}
		\caption{Mapping of the discrete joint action space to a scalar representation. The figure illustrates how a joint action from two CAVs is encoded into a unique integer ID (0-80). The highlighted example shows that if both CAVs select the action (AC, LK), the joint action is mapped to the scalar value 50.}\label{fig_action_map}
	\end{figure}

	In addition, a node is identified by the scalar value representing actions in the discrete action space. In the setting of two intelligent connected vehicles, the action-scalar mapping of the two vehicles is shown in Fig.~\ref{fig_action_map}. The action of CAV 1 (9 options, represented on the top axis) and the action of CAV 2 (9 options, on the left axis) form the 9$\times$9 grid of 81 possible joint actions. Each agent's action is a tuple of ($a_{\text {lon}}$, $a_{\text {lat}}$), where $a_{\text {lon}} \in \{-1, 0, 1\}$ corresponds to $\{\text{DC}, \text{SK}, \text{AC}\}$, and $ a_{\text {lat}} \in \{-1, 0, 1\}$ corresponds to $\{\text{LC}, \text{LK}, \text{RC}\}$. The acceleration values for discrete acceleration and deceleration actions are listed in Table~\ref{agentparameter}. The highlighted action $\bm{a}$ is mapped to the scalar value 50, denoted as $\left|\bm{a}\right|$ = 50.
	
	\textbf{Selection}: 
	The action selection in each state requires numbers of rollouts. The initial state of each rollout is the current state of the traffic. In the process of rollout, a new search tree will be generated. It should be noted that part of the tree may have been generated in the previous rollout and can be reused.
	
	Among them, the UCB calculation method is 
	\begin{equation}
		u=Q+c_{\text{puct}} \cdot p(s,\bm{a}) \sqrt{\frac{\log{n_p}}{(1+n)}},
		\label{eq_puct}
	\end{equation}
	where $n_p$ is the visit time of the parent node. On the first visit, the node's $Q$ value is assigned by action rewards. As the number of visits to the node increases, the second term gradually decreases, increasing the likelihood of visiting other nodes. Therefore, this selection strategy balances exploitation and exploration. 
	
	\begin{algorithm}[h]
		\caption{Node Expansion.}\label{alg_node_expansion}
		\begin{algorithmic}[1]
			\STATE {\textsc{ExpandNode}}$(N\{\bm{a}, Q, n\})$
			\STATE \hspace{0.5cm} \textbf{for} $\bm{a}_k$ in $\{ \bm{a} \}_{\text{legal}}$
			\STATE \hspace{1.0cm} \textbf{if} $EE$ \textbf{then}
			\STATE \hspace{1.5cm} $Q_k = \tilde{r}(\bm{a}_k)$, as defined in Equation~(\ref{eq_exp_pref})
			\STATE \hspace{1.5cm} $C = C \cup \{ N\left( \bm{a}_k, Q_k, 0 \right) \}$
			\STATE \hspace{1.0cm} \textbf{else} $C = C \cup \{ N\left( \bm{a}_k, 1, 0 \right) \}$
			\STATE \hspace{1.0cm} \textbf{end if}
			\STATE \hspace{0.5cm} \textbf{end for}
			\STATE \hspace{0.5cm} Assign prior prob for $C$ with Equation~(\ref{eq_prior_p})
		\end{algorithmic}
	\end{algorithm}
	
	\textbf{Expansion}: 
	During node expansion, we first prune actions that violate physical constraints (e.g., no right turn from the rightmost lane) or exceed vehicle velocity limits. The set of remaining valid actions is denoted by $\{ \bm{a} \}_{\text{legal}}$. For each newly expanded child node, its parent is the node from which it was expanded, and its statistics $n$, $u$, $\mathrm{Re}$, and $\mathrm{Rt}$ are initialized to zero. We employ two different methods for initializing the $Q$-value and prior probabilities of these new nodes.
	
	\begin{enumerate}
		\item \textbf{Direct expansion}: Initializes the prior $p$ to 1;
		\item \textbf{Node expansion with experiential action preference}: Compared with the PUCT algorithm in \cite{Rosin-47}, there is no prior probability information of the node. Here, we simply constructs a value mapping for the actions in the action space, so that the \textit{most promising} action can be preferentially explored. The specific mapping method is
		\begin{equation}
			\tilde{r}(\bm{a})=w_{1} R_{\text {speed }}+w_{4} P_{L C}.
			\label{eq_exp_pref}
		\end{equation}
		Compared to the full reward function in Equation (\ref{eq_rewardfun}), the absence of $R_{\text {intention}}$ and $P_{\text {collision}}$ here is because we cannot know the result with the action itself before it is excuted. Certainly, in the simulation environment, the actual reward value of the action can be obtained, but this will greatly reduce the search effenciency. The prior probability 
		\begin{equation}
			p(\bm{a}) = \frac{\tilde{r}(\bm{a})}{\sum_{\bm{b} \in \mathcal{A}_{\text{legal}}} \tilde{r}(\bm{b})}.
			\label{eq_prior_p}
		\end{equation}
		In essence, this expansion method encodes a preference for actions likely to yield higher immediate rewards, thereby guiding the search towards maximizing the $Q$-value.
	\end{enumerate} 
	
	The pseudocode for both expansion methods is presented in Algorithm~\ref{alg_node_expansion}. In the pseudocode, $\langle P, p, C, n, u, \mathrm{Re}, \mathrm{Rt}, Q\rangle$ represents the attributes of the node to be expanded, as defined in Equation~(\ref{eq_node_attr}). The set of child nodes, $C$, is initialized as an empty set. $EE$ is a boolean hyperparameter that controls the expansion strategy: if $EE$ is true, the algorithm uses the expansion method with experiential action preference.
	
	\begin{figure}[h]
		\centering
		\includegraphics[width=0.7\linewidth]{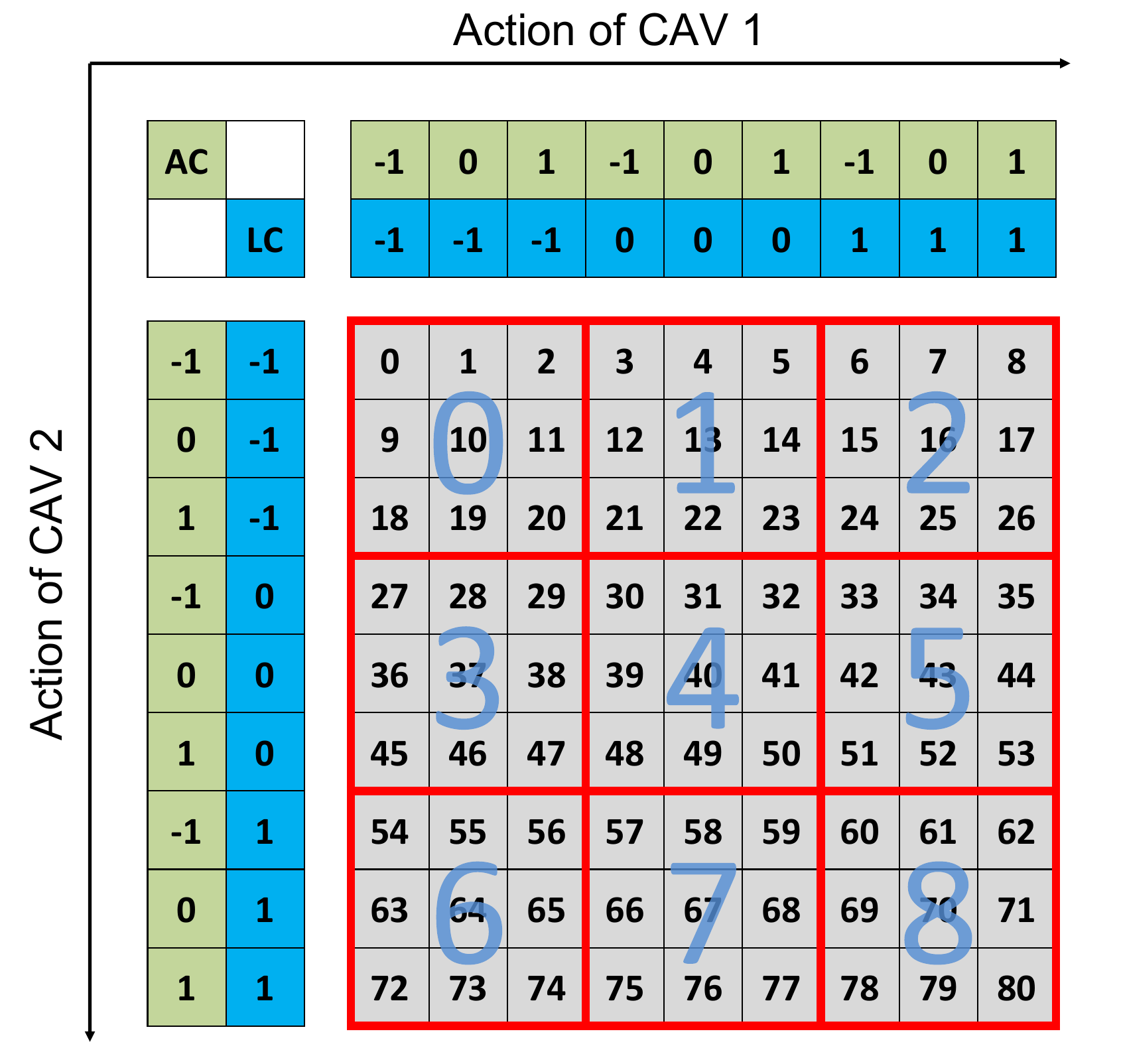}
		\caption{Grouping of the joint action space for the parallel update mechanism.}\label{fig_action_group}
	\end{figure}
	
	\textbf{Parallel update}: 
	We divide the actions in the joint action space into 9 groups, as shown in Fig.~\ref{fig_action_group}. For CAV 1, if action $\bm{a}$ is dangerous, then the set of nodes for parallel updates is 
	\begin{equation}
		G_{\text{para}}(\bm{a}) = \bigcup_{i \in \mathbb{I}_{LC}(\bm{a})} G_{i},
	\end{equation}
	where 
	\begin{equation}
		\mathbb{I}_{LC}(\bm{a}) = \{3i+\operatorname{int}\left(\bmod \left(|\bm{a}|, 9\right), 3\right) \mid i=0,1,2\}.
	\end{equation}
	
	Similarly, it is not difficult to derive the parallel update set for dangerous actions for CAV 2.
	For example of a parallel update, we assume that at the current time step, action \textbf{20} is dangerous for CAV 1. According to \textit{Definition}~\ref{def_simi_action}, all actions in group 0, 3, 6 are parallel actions of \textbf{20}, and all these 27 actions should be parallel updated. But in implementation we also exclude action \textbf{9}, \textbf{36}, and \textbf{63}, because deceleration may still be expected to avoid collisions in such situations.
	
	\subsection{Evaluation Metrics}\label{sec_eva_metric}
	
	Five evaluation metrics are calculated:
	
	\textbf{Average traffic score (ATS.)}: 
	The superiority of cooperative driving is reflected in the efficiency and safety of global traffic flow, although such a strategy is not always optimal for a single vehicle. Therefore, we use average traffic score (ATS.) to evaluate the quality of traffic flow, which is a comprehensive index. It is calculated as
	\begin{equation}
		\text{ATS.}=\frac{1}{n_{\text{sim}}}\sum_{n=0}^{n_{\text{sim}}}\frac{1}{T} \sum_{t=0}^{T-1} r_{n,t} ,
		\label{eq_ats}
	\end{equation}
	where $T$ is the simulation steps of an episode, and $n_{\text{sim}}$ denotes the number of random scenario simulations.
	
	\textbf{Average Collision rate (Coll.\%)}: The average number of collisions in the test case, which indicates the safety of the strategy. $n_{\text {collision }}$ is the number of vehicles involved in collision.
	\begin{equation}
		\text{Coll.} = \frac{1}{n_{\text{sim}}}\sum_{n=0}^{n_{\text{sim}}} \sum_{t=0}^{T-1} n_{\text{collision},t} .
	\end{equation}
	
	\textbf{Average Arrival rate (Arri.\%)}: Average percentage of vehicles successfully arrive its destination in all test cases.
	\begin{equation}
		\text{Arri.\%} = \frac{1}{n_{\text{sim}}} \sum_{n=0}^{n_{\text{sim}}} \frac{1}{N} \sum_{t=0}^{T-1} n_{\text{arrived},t} .
	\end{equation}
	
	\textbf{Average velocity (Velo.1 and Velo.2)}: The mean value of average velocity (in $m/s$) of all vehicles per episode of all test cases.
	
	\begin{equation}
		\text{Velo.$i$} = \frac{1}{n_{\text{sim}}} \sum_{n=0}^{n_{\text{sim}}} \frac{1}{T} \sum_{t=0}^{T-1} v_{i,t} .
	\end{equation}
	
	\begin{figure}[h]
		\centering
		\includegraphics[width=0.95\linewidth]{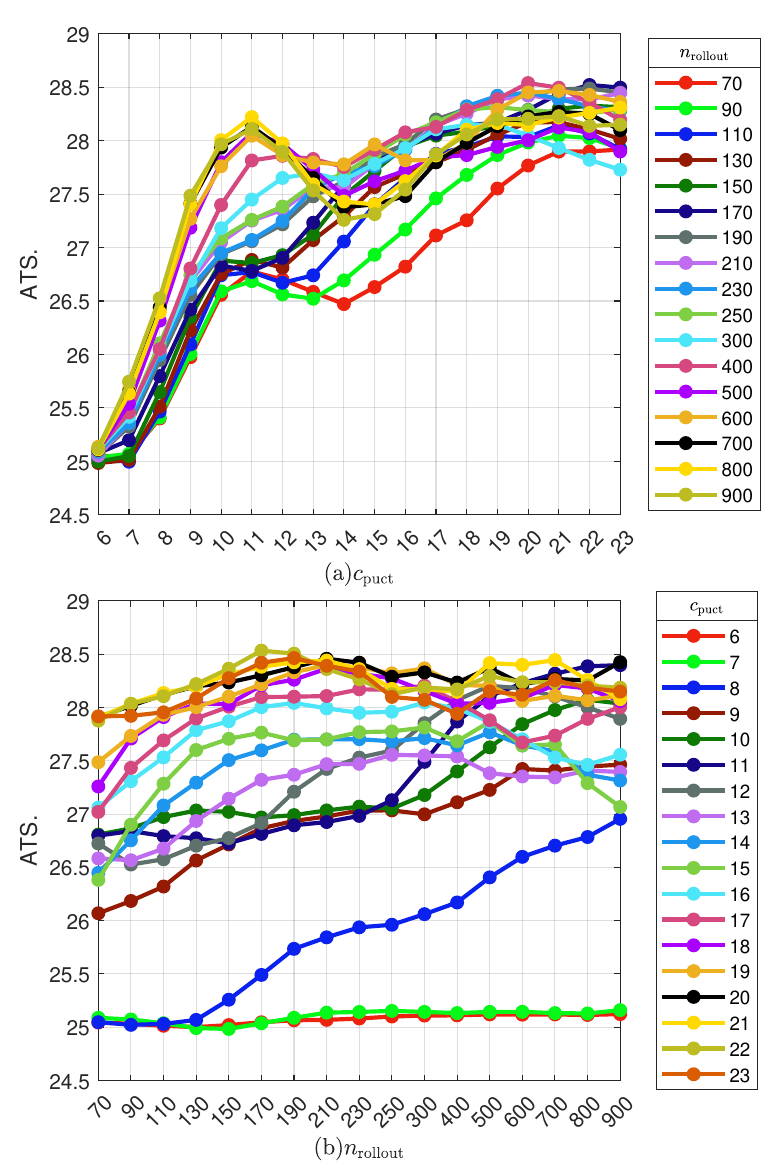}
		\caption{Sensitivity analysis of algorithm performance (ATS.) with respect to hyperparameters $c_{\text{puct}}$ and $n_{\text{rollout}}$. The plots were generated by performing orthogonal experiments across a grid of parameter values. (a) ATS as a function of the exploration constant $c_{\text{puct}}$. (b) ATS as a function of the rollout budget $n_{\text{rollout}}$. }\label{fig_puct_mesh}
	\end{figure}
	
	\subsection{Results}
	
	\subsubsection{Hyperparameter Sensitivity Analysis}\label{sec_hyper_sens}
	
	To determine the optimal configuration for our proposed MCTS algorithm, we analyzed the impact of hyperparameters on model performance. Orthogonal experiments on two key hyperparameters $c_{\text{puct}}$ and $n_{\text{rollout}}$ are conducted. It can be seen from Fig.~\ref{fig_puct_mesh}(a), under the problem settings of this paper, the \textbf{ATS}. upper bound of the MCTS algorithm is around 28.5 (without assigning a random seed).
	
	First, we analyze the impact of the exploration-exploitation trade-off, controlled by $c_{\text{puct}}$. Fig.~\ref{fig_puct_mesh}(a) plots \textbf{ATS.} as a function of $c_{\text{puct}}$ for various rollout budgets. The relationship is non-monotonic. When $c_{\text{puct}}$ is too small (e.g., $\le$10), the algorithm's exploration is insufficient, causing it to prematurely converge on locally optimal solutions and yield poor performance. Conversely, as described by Equation~(\ref{eq_puct}), an excessively large $c_{\text{puct}}$ over-encourages exploration, leading to a shallow search depth that fails to identify actions with high long-term rewards. The plot confirms this, showing that the best performance is consistently achieved within the range $c_{\text{puct}} \in [17, 22]$.
	
	The influence of the computational budget $n_{\text{rollout}}$ is shown in Fig.~\ref{fig_puct_mesh}(b). For well-tuned $c_{\text{puct}}$ values (e.g., the black and yellow curves), performance improves significantly as $n_{\text{rollout}}$ increases from 70 to approximately 200. Beyond this point, the performance gains become marginal while the computational cost continues to increase linearly. This indicates that a budget of around 200 rollouts is sufficient for the agent to make robust decisions in our problem setting.
	
	To balance high performance with computational efficiency, we select hyperparameter values of $c_{\text{puct}} = \text{21}$ and $n_{\text{rollout}} = \text{200}$. It is worth noting that this optimal range is specific to our defined problem and reward structure; different settings may require re-tuning.
	
\begin{figure}[htbp]
		\centering
		\includegraphics[width=0.95\linewidth]{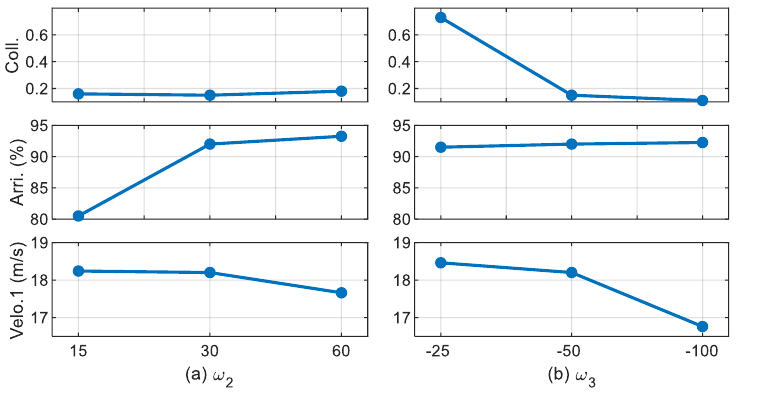}
		\caption{Impact of varying the intention reward ($w_2$) in (a) and the collision penalty ($w_3$) in (b) on key performance metrics.}\label{fig_reward_sensitivity}
	\end{figure}

	To empirically validate the robustness of the reward function weights, we conducted a sensitivity analysis on the intention reward ($w_2$) and the collision penalty ($w_3$). We individually adjusted either $w_2$ or $w_3$ while holding all other parameters constant. This method tests the relative importance of the adjusted parameter against the fixed rewards for other objectives (e.g., efficiency). The results are visualized in Fig.~\ref{fig_reward_sensitivity}.
	
	As shown in Fig.~\ref{fig_reward_sensitivity}(a), decreasing $w_2$ to 15 significantly weakens the agent's motivation for its primary mission, causing the arrival rate to plummet from 92.0\% to 80.5\%. Conversely, while increasing $w_2$ to 60 further boosts the arrival rate to 93.25\%, this marginal gain comes at the cost of safety.
	This demonstrates that $w_2=30$ is a well-calibrated value, ensuring high task completion without encouraging reckless behavior. In Fig.~\ref{fig_reward_sensitivity}(b), as we relax the collision penalty to -25, the collision rate skyrockets from 0.15 to 0.73, indicating an insufficient deterrent against hazardous actions. On the other hand, intensifying the penalty to -100 makes the agent overly conservative. While this further reduces collisions, it significantly curtails the vehicle's average velocity, impairing overall traffic efficiency.
	
	The analysis above confirms that the chosen parameters ($w_2=30$, $w_3=-50$) achieve a well trade-off between the competing objectives of safety, mission success, and traffic efficiency.
	
	\begin{figure*}[h]
		\centering
		\includegraphics[width=0.95\linewidth]{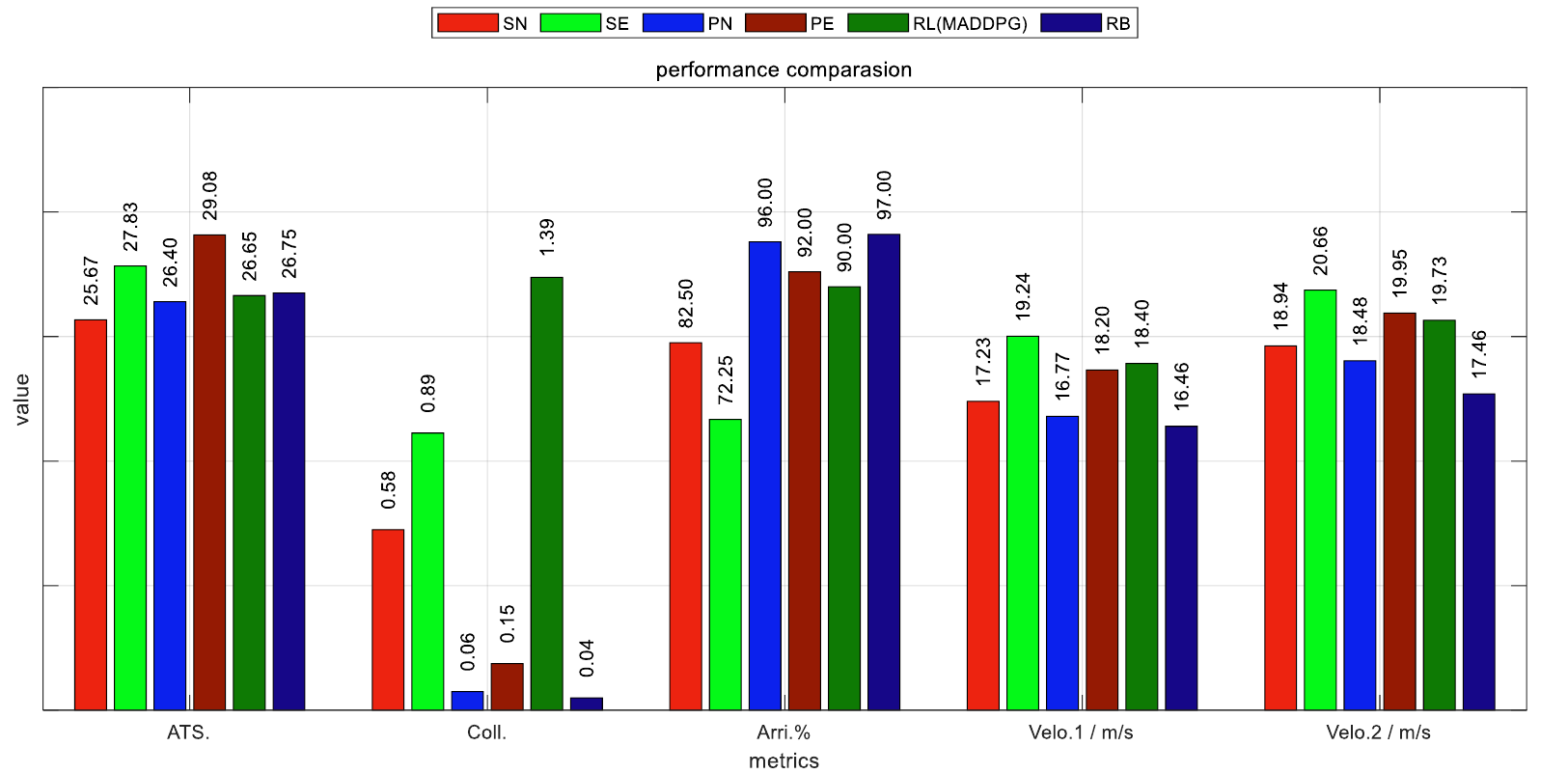}
		\caption{
			Comparative performance of the proposed MCTS variants and baseline methods across 5 key metrics. The results demonstrate the superiority of our algorithm (PE), which achieves the highest overall efficiency while maintaining a near-perfect safety record. The ablation study highlights the synergy of our contributions: the Parallel Update (PN) ensures safety, while experiential action preference (SE) boosts speed, and their combination in PE yields the best-balanced performance.
		}\label{fig_performance_hist}
	\end{figure*}
	
	\subsubsection{Performance Comparasion}\label{sec_perf_comp}
	
	To rigorously evaluate our proposed method, we conducted a comparative analysis against a suite of baseline algorithms. The methods include our full model with both parallel update and experiencial action preference (PE), and its two ablations: parallel update only (PN) and action preference only (SE). We also compare against a Standard MCTS (SN), a well trained multi-agent reinforcement learning algorithm (RL-MADDPG), and a Rule-Based (RB) approach based on the model described in Section~\ref{sec_world_model}. Performance was assessed across 5 key metrics defined in Section~\ref{sec_eva_metric}.
	
	As shown in Fig.~\ref{fig_performance_hist}, rule-based algorithms meet the basic requirements for safely completing tasks, performing well in both the collision and arrival rate metrics (\textbf{Coll.} and \textbf{Arri.\%}). However, they fall behind other methods in terms of speed (\textbf{Velo.1} and \textbf{Velo.2}). Total of 4 collisions and less than 100\% arrival rate observed in the rule-based method are due to unavoidable obstacles created by randomly generated traffic flow. RL algorithm performs slightly worse overall compared to rule-based methods. Although they achieve higher speeds, they fail to handle dynamic and unpredictable environments, leading to a higher number of collisions and lower performance in \textbf{Arri.\%}. This results in suboptimal overall performance (\textbf{ATS.}). It should be noted that the driving strategy of the RL algorithm is aggressive under the reward function design used in this paper.
	
	The basic MCTS algorithm performs slightly worse than the RB and RL algorithms in terms of \textbf{ATS.} and \textbf{Arri.\%}, but it shows significant improvement in \textbf{Coll.} compared to the RL algorithm, although it lacks an advantage in speed. After incorporating the parallel update, the MCTS algorithm demonstrates superior performance in \textbf{Coll.}—only 6 collisions occurred across all random scenario tests—though there is a slight decrease in speed.
	
	\begin{figure}[h]
		\centering
		\includegraphics[width=1.0\linewidth]{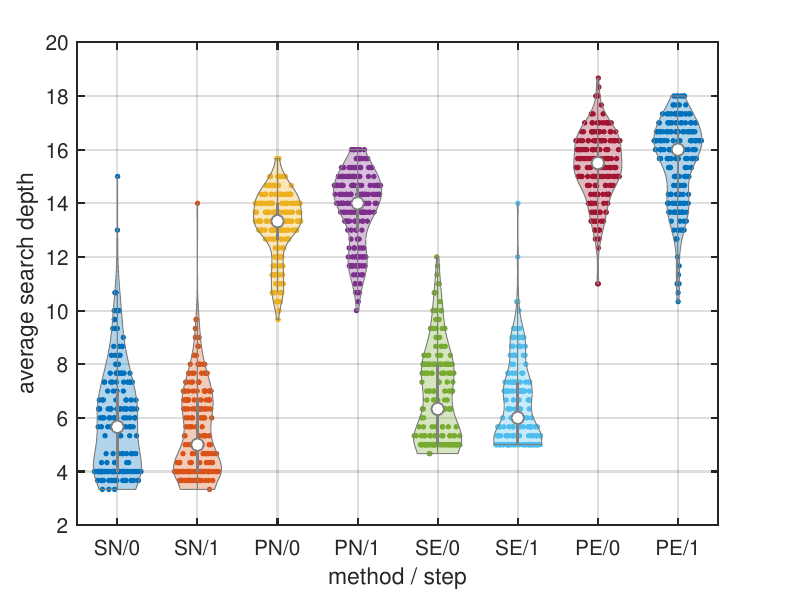}
		\caption{
			Violin plots comparing the average MCTS search depth for each method at the first 2 steps (/0 and /1) within a fixed 200-rollout budget.
		}\label{fig_search_dep_vln}
	\end{figure}
	
	When action preferences are implemented, the \textbf{ATS.} increases by 8.37\%, which, as shown in Fig.~\ref{fig_performance_hist}, is primarily due to a significant increase in vehicle speed. This is because the rule in Equation (\ref{eq_prior_p}) leads the agent to explore acceleration actions more frequently, ultimately converging to those actions with higher probability. However, the increased speed also results in a more aggressive driving style, leading to poor performance in \textbf{Coll.} and \textbf{Arri.\%}.
	
	The MCTS algorithm with both parallel updates and action preferences performs optimally overall, showing good results across all metrics. The parallel update method makes the algorithm more sensitive to hazardous actions, improving safety performance. After quickly excluding dangerous actions, the algorithm can more fully explore actions with greater potential, allowing the agent to execute more rational actions over longer horizon. Compared to the RL algorithm, the number of collisions is reduced by 89.21\%, traffic efficiency is improved by 12.40\% over the rule-based method, and the overall task completion rate remains above 90\%.
	
	It should be noted that \textbf{Velo.2} is higher than \textbf{Velo.1} in all cases, it is because the destination of CAV 2 is further away from the starting point, and there are more opportunities to excute accelerating actions.
	
	\subsubsection{Search Depth Statistics}
	
	Exploration depth reflects an agent's ability to anticipate future consequences of current actions. A greater search depth within a fixed computational budget often enables an agent to discover sophisticated, long-term strategies that surpass myopic decision-making. To provide a mechanistic explanation for the performance results in Section~\ref{sec_perf_comp}, we statistically analyzed the search depth achieved by each MCTS variant.
	
	\begin{figure*}[ht]
		\centering
		\includegraphics[width=1.0\linewidth]{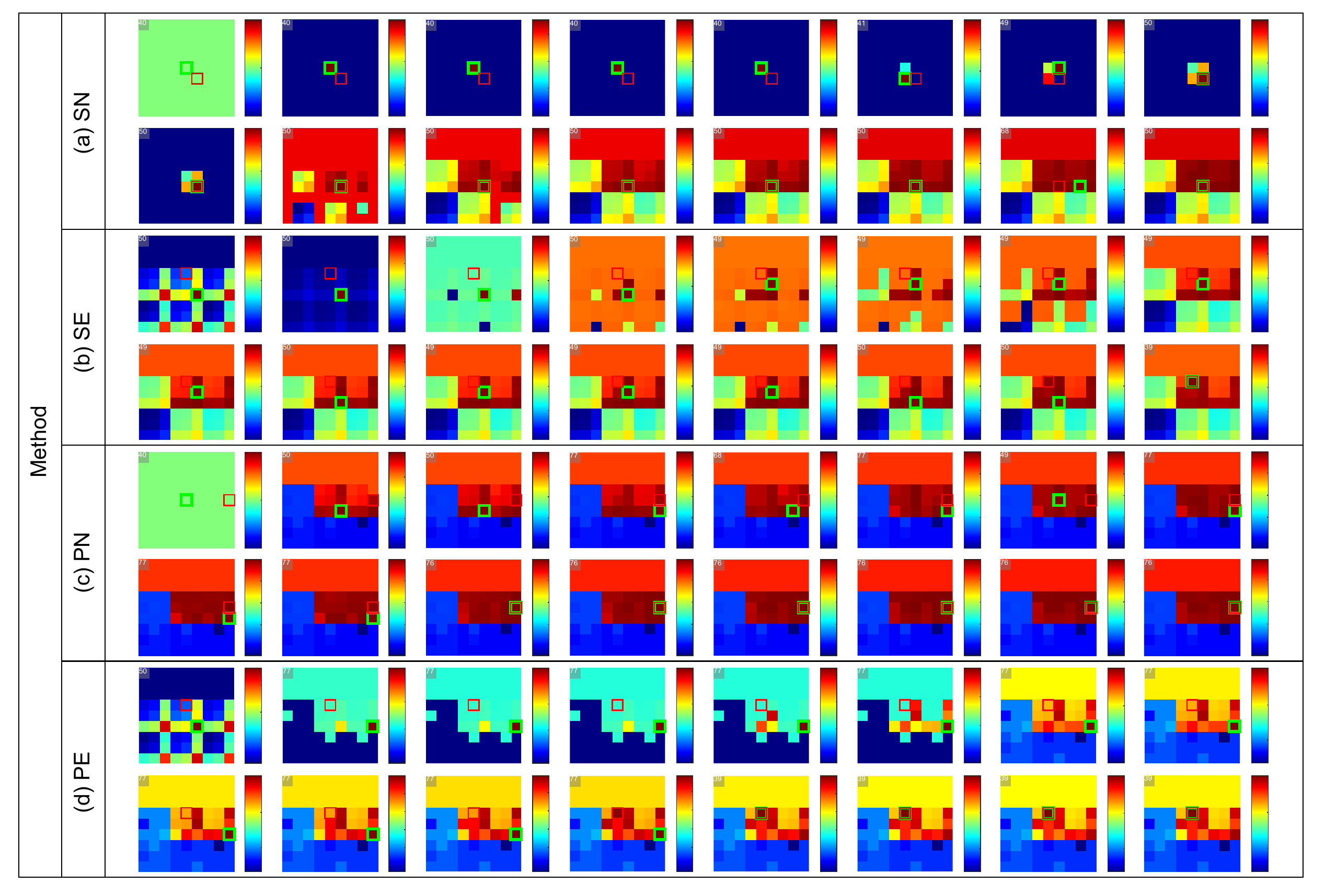}
		\caption{Rollout detail of the first step, comparing the search efficiency of 4 algorithm configurations. The figure shows the evolution of the Q-value distribution over the joint action space during a single decision step. Each row represents a different method as defined in Section~\ref{sec_perf_comp}. Each row consists of 16 snapshots taken at even intervals during a 200-rollout search.}\label{fig_q_heat}
	\end{figure*}
	
	The results are presented as violin plots in Figure~\ref{fig_search_dep_vln}. We compare the 4 MCTS methods at the first 2 steps labeled '/0' and '/1'. The most striking result is the significant increase in average search depth for methods incorporating the parallel update mechanism (PN and PE) compared to those without (SN and SE). This provides direct evidence for our hypothesis: by rapidly pruning large regions of unsafe actions, the parallel update conserves the rollout budget, allowing the search to penetrate much deeper down the promising branches of the game tree. Instead of wasting evaluations on dead-end paths (i.e., collisions), the algorithm focus on longer-term planning, more than doubling the average search depth.
	
	In contrast, Action Preference alone does not guarantee a deeper search and can even lead to shallower exploration by focusing on immediately rewarding, but potentially simple, actions. However, the synergistic effect in PE MCTS is clear. It achieves the highest search depth, as the Action Preference guides the search towards promising areas, while the parallel update efficiently secures these paths by eliminating surrounding threats. This combination enables a confident and deep exploration of high-potential strategies.
	
	While deeper search is beneficial, we acknowledge its trade-offs, such as higher computational demands and potential sensitivity to prediction errors in noisy environments, which remain important considerations for real-world deployment.
	
	\subsubsection{Search Efficiency Comparasion}
	
	To provide an intuitive and qualitative analysis of our algorithm's search efficiency, we visualize the internal dynamics of the MCTS search process. As shown in Fig.~\ref{fig_q_heat}, for 4 different MCTS algorithm configurations as defined in Section~\ref{sec_perf_comp}, we visualized the action's relative Q-value at evenly spaced intervals during a single step's rollouts under the same traffic condition.
	
	We specifically select the decision at time step 0 for this analysis. This choice is deliberate: as all vehicles are randomly placed in the first half of the road, the initial state features the most intense right-of-way conflicts, making early decisions critical. Furthermore, analyzing from a guaranteed identical starting point ensures a fair and direct comparison of the algorithms' search patterns.
	
	\begin{figure*}[h]
		\centering
		\includegraphics[width=1.0\linewidth]{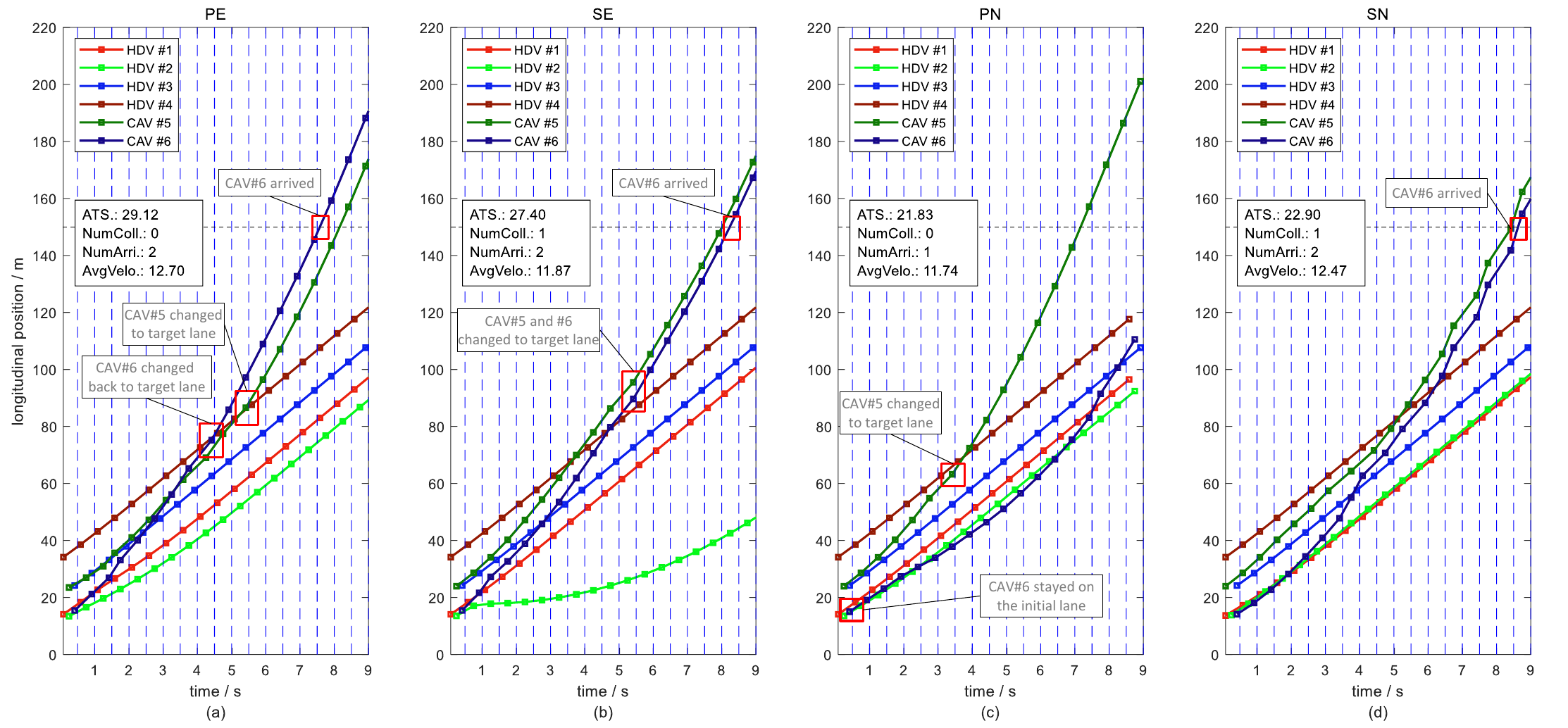}
		\caption{Time-space diagrams of the cooperative driving case, comparing the performance of four MCTS methods. Each subplot displays the trajectories of 6 vehicles over 9 second for a specific method. The region between each pair of blue dashed lines can be considered a snapshot of the traffic flow at a single time step; the horizontal offset of each point is used to distinguish the vehicle's lane. In this scenario, CAV 6 is the primary CAV attempting to exit, and CAV 5 is the assisting CAV.}\label{fig_sp_case_1_ts}
	\end{figure*}
	
	In Fig.~\ref{fig_q_heat}, we show 16 equally intervaled rollout results of 200 rollouts in the first step of each method. Each heatmap represents the 81-action joint action space. Within each heatmap, Color corresponds to the relative Q-value (blue is low, yellow is high), the green box highlights the action with the highest Q-value at that snapshot, and the red box indicates the final action selected after the entire search is complete. The maximum $Q$ value of the current rollout is marked in the upper left corner of the heat map. It should be noted that the color mapping setting of each heat map is independent, indicating only the relative relationship of the $Q$ value.
	
	We begin by observing the baseline Standard MCTS (SN) in Fig.~\ref{fig_q_heat}(a). The search process is inefficient, initially exploring a wide range of unsafe, low-value actions (indicated by the large dark blue areas in the early snapshots. In fact, all actions during the first 18 rollouts led to collisions). This sequential and unguided exploration wastes a significant portion of the computational budget before discovering promising candidate actions, heightening the risk of converging on a suboptimal choice.
	
	The impact of incorporating experiential action preference is immediately evident when comparing Fig.~\ref{fig_q_heat}(a) with Fig.~\ref{fig_q_heat}(b), and Fig.~\ref{fig_q_heat}(c) with Fig.~\ref{fig_q_heat}(d). In Fig.~\ref{fig_q_heat}(b) and Fig.~\ref{fig_q_heat}(d), the very first snapshot shows that the search is immediately biased towards a promising, high-value region. This mechanism effectively provides a good start, steering the search away from obviously poor choices and towards areas of the action space with greater potential from the outset.
	
	The key advantage of our Parallel Update method is vividly demonstrated in Fig.~\ref{fig_q_heat}(c) and Fig.~\ref{fig_q_heat}(d). Observe how, after just a few rollouts, large, contiguous areas of the action space turn dark blue simultaneously. This visualizes our parallel backpropagation, where exploring a single unsafe action allows the algorithm to collectively prune an entire region of related, unsafe actions. This ability to eliminate patches of the search space dramatically accelerates the pruning of unpromising branches.
	
	The synergy of both mechanisms is most apparent in our full model, PE, shown in Fig.~\ref{fig_q_heat}(d). It combines the focused start of experiencial action preference with the rapid, large-scale pruning of parallel update. As a result, the search quickly dismisses vast, suboptimal regions and concentrates almost the entire computational budget on refining the choice between a few excellent candidate actions. This heightened search efficiency is critical for real-time decision-making, enabling the agent to find more robust and intelligent solutions within a limited time horizon.
	
	\begin{figure}[h]
		\centering
		\includegraphics[width=0.95\linewidth]{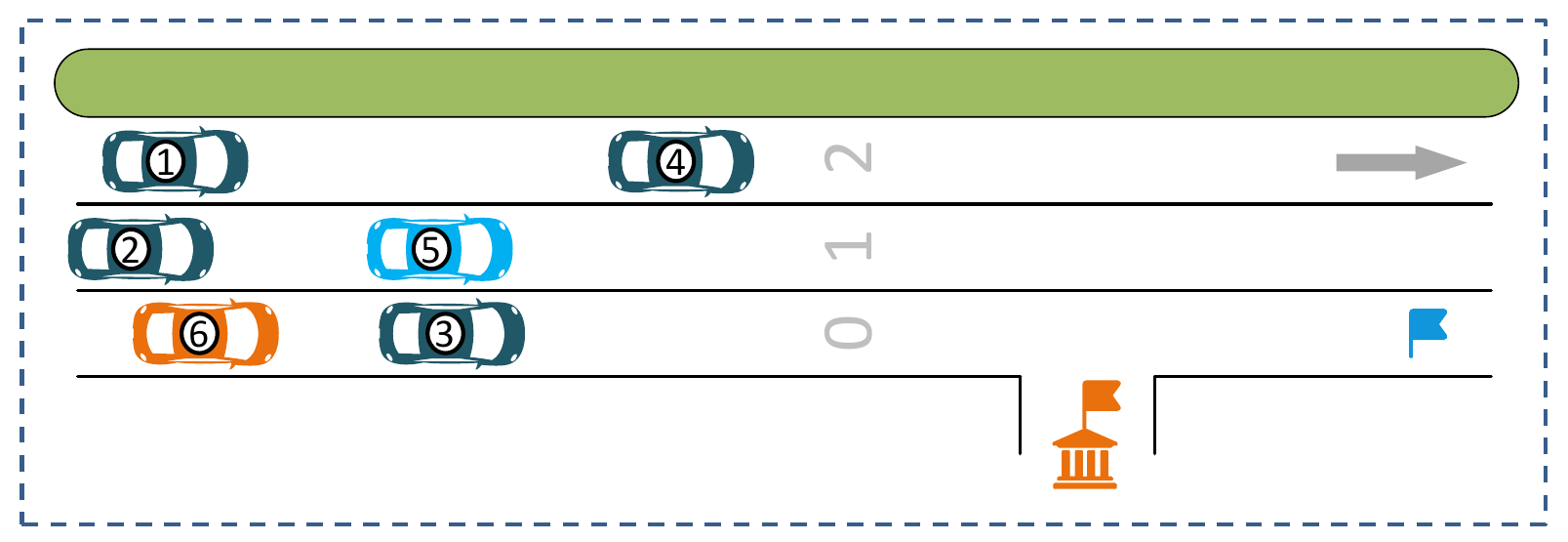}
		\caption{Initial state of a challenging cooperative driving case. CAV 6 is initially boxed in by several HDVs. Our algorithm enables CAV 5 to proactively create a gap for CAV 6 to merge, demonstrating a sophisticated, multi-agent strategy.}\label{fig_sp_case_1}
	\end{figure}
	
	\subsubsection{Typical Cooperative Case}
	
	Through analysis of the simulation results from the randomly generated experiments detailed in Section~\ref{sec_case_setting}, we identified some particularly challenging instances where the algorithm discovered strategies demonstrating an emergent intelligence superior to that of typical, myopic human-like driving.
	
	As depicted in in Fig.~\ref{fig_sp_case_1}, the initial state presents a significant challenge. The CAV 6, intending to take the upcoming exit, is effectively 'boxed in' by a formation of HDV 2 and 3. Its path is blocked, and the presence of HDV 1 in the adjacent lane prevents any immediate lane-changing or overtaking maneuvers. Vehicles numbered 1 through 6 have initial longitudinal positions of [14.10, 13.42, 24.32, 34.15, 23.52, 15.47]~m and corresponding lane indices of [2, 1, 0, 2, 1, 0], respectively.
	
	In this situation, a conventional algorithm or a myopic agent controlling the other CAV 5 would likely prioritize its own progress by maintaining speed or accelerating. This is observed in our baseline comparisons, as shown in the time-space diagrams in Fig.~\ref{fig_sp_case_1_ts}(b), (c), and (d). In these instances, CAV 5 acts greedily, leaving CAV 6 trapped behind the HDV formation, which severely compromises its mission and the overall traffic efficiency.
	
	In contrast, our proposed algorithm discovers a highly sophisticated and non-intuitive cooperative strategy, as illustrated in Fig.~\ref{fig_sp_case_1_ts}(a). At the critical first step, the algorithm directs CAV 5 to perform a sacrificial deceleration. While locally suboptimal for CAV 5 itself, this maneuver creates a subtle ripple effect through the traffic stream. It compels the trailing HDV 2, whose behavior is governed by a conservative driving model, to also slow down in response. This coordinated action opens a sufficient and safe gap in the middle lane. CAV 6 then seizes this fleeting opportunity to execute a smooth lane change. Once clear of the obstruction, CAV 6 has ample space to accelerate towards its destination, ultimately achieving its goal far more efficiently than in the baseline scenarios. 
	
	By identifying a globally optimal solution over myopic individual gains, our algorithm orchestrates a complex maneuver that a typical human driver or a greedy algorithm would likely overlook. The resulting improvement in traffic efficiency and mission success showcases the profound potential of our method for navigating complex, multi-agent interactions.
	
	\section{Conclusion}\label{sec_Conclusion}
	
	This proposed a value-based parallel update MCTS method, whose core innovation is a parallel update mechanism based on action similarity, enabling the algorithm to search efficiently within the vast joint action space. Through extensive experiments in mixed-traffic simulation environment, we validated the effectiveness of our method. The results show that, compared to standard MCTS and other mainstream baseline methods, our algorithm demonstrates superior performance in significantly enhancing traffic safety and operational efficiency (improving the average traffic score). An in-depth analysis of the algorithm's internal search dynamics further confirmed that the parallel update mechanism achieves deeper, more focused exploration by effectively pruning unsafe regions, thereby enabling the algorithm to discover and execute complex, non-myopic, and globally optimal cooperative strategies.
	
	Despite these promising results, it is important to acknowledge the limitations of the current study, which in turn motivate the directions for future work. Our framework operates under several simplifying assumptions. First, it relies on a centralized coordinator with ideal, error-free V2X communication, which may not hold in real-world deployments subject to latency and packet loss. Second, the HDV behaviors are simulated using homogeneous, rule-based models; while providing a consistent testbed, they do not capture the full spectrum of heterogeneous and potentially irrational human driving styles. Furthermore, the study is conducted in a specific, relatively constrained highway merging scenario with a small number of vehicles and a discrete action space. The scalability and generalizability of our approach to more complex urban environments, higher traffic densities, and continuous control remain to be demonstrated. Finally, while our method empirically improves safety, it does not provide formal safety guarantees, a critical requirement for deployment. The emergence of sacrificial strategies also highlights the need for an explicit ethical framework to govern such decisions, a dimension this work identifies but does not resolve.
	
	In the future, this work could be extended as follows.
	Firstly, the current action similarity metric, while effective due to its basis in TTC, is fundamentally static. A significant future direction involves developing dynamic, state-aware action abstraction capabilities. By incorporating DRL methods, it is possible to learn a function that dynamically outputs action similarity scores based on the current global traffic state, thus making the parallel update mechanism more intelligent and adaptive. Secondly, to advance toward real-world applicability, the current centralized framework should be extended. Future research will focus on developing a decentralized MCTS algorithm robust to communication uncertainties like latency and packet loss. Thirdly, a powerful extension is to create a hybrid system that combines our MCTS planner with deep neural networks. A neural network trained to predict high-quality policies and state values can provide strong priors, guiding the MCTS to search more efficiently and unlocking a higher level of performance and adaptability.
	

	\bibliographystyle{IEEEtran.bst}
	\bibliography{ref/ref.bib}
	
	\vfill

\end{document}